\documentclass[a4paper,fleqn,usenatbib]{mnras}

\usepackage{mathptmx}

\usepackage{ae,aecompl}

\usepackage{graphicx}	
\usepackage{amsmath}	
\usepackage{amssymb}	

\usepackage{color,epsfig} 
\usepackage[utf8]{inputenc} 

\usepackage{multirow}

\usepackage[flushleft]{threeparttable}
\makeatletter
\newlength{\abovecaptionskip}%
\makeatother

\newcommand{\be}{\begin{equation}}
\newcommand{\ee}{\end{equation}}

\newcommand*\diff{\mathop{}\!\mathrm{d}}
\newcommand{\vect}[1]{\mathbf{#1}}

\newcommand{\rtdis}{R^{\rm dis}_{\rm t}}

\newcommand{\rtsep}{R^{\rm sep}_{\rm t}}

\newcommand{\mstar}{M_{\star}}
\newcommand{\rstar}{R_{\star}}

\newcommand{\rp}{R_{\rm p}}
\newcommand{\thetaint}{\theta_{\rm int}}
\newcommand{\thetacol}{\theta_{\rm col}}
\newcommand{\thetaintbar}{\bar{\theta}_{\rm int}}
\newcommand{\thetacolbar}{\bar{\theta}_{\rm col}}
\newcommand{\thetacolbarrel}{\bar{\theta}^{\rm rel}_{\rm col}}
\newcommand{\psirel}{\psi_{\rm rel}}
\newcommand{\rcol}{R_{\rm col}}

\newcommand{\emax}{e^{\rm max}}
\newcommand{\emin}{e^{\rm min}}

\newcommand{\dt}{\Delta t}
\newcommand{\dtnc}{\Delta t_{\rm nc}}
\newcommand{\dtmb}{\Delta t_{\rm mb}}
\newcommand{\dtheta}{\Delta \theta}

\newcommand{\dthetanc}{\Delta \theta_{\rm nc}}

\newcommand{\dthetamb}{\Delta \theta_{\rm mb}}

\newcommand{\mh}{M_{\rm h}}

\newcommand{\tmin}{t_{\rm min}}
\newcommand{\amin}{a_{\rm min}}

\newcommand{\ex}{\vect{e}_{\rm x}}
\newcommand{\ey}{\vect{e}_{\rm y}}
\newcommand{\ez}{\vect{e}_{\rm z}}

\newcommand{\epara}{\vect{e}_{\parallel}}

\newcommand{\er}{\vect{e}_{\rm r}}

\newcommand{\ur}{\vect{u_{\rm r}}}
\newcommand{\ut}{\vect{u_{\rm t}}}
\newcommand{\dr}{\Delta \vect{R}}
\newcommand{\dv}{\Delta \vect{v}}
\newcommand{\ddj}{\Delta \vect{j}}
\newcommand{\de}{\Delta \vect{e}}

\newcommand{\pcol}{P_{\rm col}}
\newcommand{\tcol}{t_{\rm col}}

\newcommand{\gammamu}{\gamma_{\rm mu}}

\newcommand{\domegad}{\Delta \omega_{\rm d}}
\newcommand{\domegan}{\Delta \omega_{\rm n}}

\newcommand{\dthetarel}{\Delta \theta_{\rm rel}}
\newcommand{\thetacolrel}{\theta_{\rm col}^{\rm rel}}

\def\m6{\, M_6}
\def\ms{\, m_{\star}}
\def\rs{\, r_{\star}}
\def\a3{\, a_3}

\def\d{\, \mathrm{d}}
\def\msun{\, \mathrm{M}_{\hbox{$\odot$}}}
\def\rsun{\, \mathrm{R}_{\hbox{$\odot$}}}
\def\gcm3{\, \rm g \, cm^{-3}}

\def\kpc{\, \rm kpc}
\def\yr{\, \rm yr}

\def\au{\, \rm{au}}
\def\erg{\, \rm erg}
\def\ergs{\, \rm erg\, s^{-1}}

\title[Streams collision in double TDEs]{Streams collision as possible precursor of double tidal disruption events}

\author[Clément Bonnerot and Elena M. Rossi]{Clément Bonnerot$^{1,2}$\thanks{E-mail: bonnerot@tapir.caltech.edu}
and Elena M. Rossi$^{1}$
\\
$^{1}$Leiden Observatory, Leiden University, PO Box 9513, 2300 RA, Leiden, the Netherlands\\
$^{2}$TAPIR, Mailcode 350-17, California Institute of Technology, Pasadena, CA 91125, USA
}

\date{Accepted XXX. Received YYY; in original form ZZZ}

\pubyear{2017}

\begin{document}
\label{firstpage}
\pagerange{\pageref{firstpage}--\pageref{lastpage}}
\maketitle

\begin{abstract}

The rate of tidal disruption events (TDEs) can vary by orders of magnitude depending on the environment and the mechanism that launches the stars towards the black hole's vicinity. For the largest rates, two disruptions can take place shortly one after the other in a double TDE. In this case, the two debris streams may collide with each other before falling back to the black hole resulting in an electromagnetic emission that is absent from single TDEs. We analytically evaluate the conditions for this streams collision to occur. It requires that the difference in pericenter location between the two disruptions makes up for the time delay between them. In addition, the width of the streams must compensate for the vertical offset induced by the inclination of their orbital planes. If the double TDE happens following the tidal separation of a binary, we find that the streams can collide with a probability as high as $44\%$. We validate our analytical conditions for streams collision through hydrodynamical simulations and find that the associated shocks heat the gas significantly. If photons are able to rapidly escape, a burst of radiation ensues lasting a few days with a luminosity $\sim 10^{43} \ergs$, most likely in the optical band. This signal represents a precursor to the main flare of TDEs that could in particular be exploited to determine the efficiency of disc formation from the stellar debris.

\end{abstract}

\begin{keywords}
black hole physics -- hydrodynamics -- galaxies: nuclei.
\end{keywords}



\section{Introduction}
\label{introduction}

A tidal disruption event (TDE) happens when a star gets so close to a supermassive black hole that it is destroyed by the strong tidal forces of the compact object. Following the disruption, the stellar debris evolves to form a thin and elongated stream that keeps orbiting the black hole. Roughly half of the gas within that stream is bound on highly eccentric orbits while the rest gets unbound and escapes on hyperbolic trajectories \citep{lacy1982,rees1988,phinney1989,evans1989}. As the bound part of the stream comes back to the disruption site, it undergoes complex interactions during which shocks eventually lead to the formation of an accretion disc. Disc formation is primarily driven by self-crossing shocks induced by relativistic apsidal precession \citep{hayasaki2013,dai2015,shiokawa2015,bonnerot2016-circ,sadowski2016}. In addition, this process depends strongly on the black hole spin that can delay the occurrence of the first orbit crossing through Lense-Thirring precession \citep{dai2013,guillochon2015,hayasaki2016-spin} and on the gas cooling efficiency, which determines the geometry of the newly-formed disc and the amount of escaping radiation \citep{jiang2016,bonnerot2017-stream}. The theoretical understanding of this phase of evolution remains however limited by the fact that its study in the most general case is numerically challenging.

A few tens of TDE candidates have been discovered so far. They have been observed in various electromagnetic bands, in particular at optical/UV \citep{gezari2009,gezari2012,van_velzen2011,arcavi2014,holoien2014,holoien2016-14li,holoien2016-15oi,blagorodnova2017,hung2017-16axa} and X-ray wavelengths \citep{bade1996,komossa2004,komossa1999-ngc5905,esquej2008,maksym2010,saxton2017} as flares lasting from a few months to several years. While the X-ray component almost certainly comes from gas accreting onto the black hole, the nature of the lower energy signal remains debated with the main possible sources being shocks from the disc formation process \mbox{\citep{lodato2012,piran2015-circ,bonnerot2017-stream}} and reprocessed accretion luminosity by a surrounding gaseous envelope \citep{loeb1997,guillochon2014-10jh,metzger2016,roth2016}. Recently, observations of variability lags and delayed emission in X-ray with respect to the lower energy optical and UV signals have been attributed to the disc formation process, providing the first observational signatures of this phase of evolution \citep{pasham2017,gezari2017-15oi}.

However, one major issue when attempting to constrain theoretical models with observational data relates to the lack of emission prior to the onset of disc formation. Several mechanisms have been proposed that produce earlier radiation, but they appear to be either too dim or too short-lived to be easily detected. A faint optical flare can for example result from the recombination of hydrogen within the tidal stream \citep{kasen2010}. Alternatively, radiation can emerge from the strong stellar compression happening at pericenter if the star is disrupted on a deeply plunging trajectory. In this situation, an X-ray shock breakout signal can be emitted but the associated burst has a duration of only a few tens of seconds \citep{kobayashi2004,brassart2008,guillochon2009,brassart2010}. In addition, nuclear reactions can be triggered by the compression whose radioactive output results in an optical flare upon reprocessing by the expanding gas distribution, a phenomenon especially promising for white dwarf disruptions \citep{rosswog2008,rosswog2009,macleod2016}. Nevertheless, even for these initially dense objects, it is still debated whether the conditions required for nuclear burning are actually met \citep{tanikawa2017}. Finally, further radiation can originate from strong relativistic precession at pericenter that results in shocks between the leading and trailing edges of an elongated white dwarf, leading to prompt gas accretion \citep{haas2012,evans2015}.

TDEs are generally thought to originate from encounters between stars surrounding the black hole that occasionally scatter one of them on a trajectory entering the tidal sphere. For this mechanism, the disruption rate per galaxy is predicted to be $\dot{N} \gtrsim 10^{-4} \yr^{-1}$ by standard two-body relaxation calculations \citep{magorrian1999,wang2004,stone2016-rates}. However, more exotic dynamical processes exist that are expected to produce a much higher rate of disruptions up to $\dot{N} \approx 1 \yr^{-1}$. For instance, if the galaxy contains a binary black hole with about a parsec separation, stars can be efficiently funnelled into the disruption radius of the primary through a combination of secular Kozai interactions and scattering by the secondary compact object \citep{chen2009,chen2011}.\footnote{The resulting TDEs would however not be affected by the presence of the secondary black hole, which only happens if the binary reaches separations smaller than around a milli-parsec \citep{coughlin2017,vigneron2018}.} A high rate of TDEs can also be caused by the presence of an eccentric nuclear disc whose stabilizing mechanism involves strong torques able to efficiently deflect stars into plunging trajectories \citep{madigan2017}. Finally, a TDE boost is expected if the stars evolve in a triaxial potential owing to the possibility of chaotic orbits \citep{merritt2004}. Some of the above mechanisms may account for the preference of optical TDEs for rare E+A galaxies \citep{french2016,french2017}.

Two TDEs can also happen shortly one after the other when a stellar binary approaches a black hole on a nearly radial orbit. In this situation, \citet{mandel2015} showed that the binary separation can be followed by the sequential tidal disruptions of the two stars and estimated that this mechanism represents around 10\% of all TDEs.\footnote{Recently, \citet{coughlin2018} showed that the same type of double disruption can occur if a stellar binary encounters a binary black hole.} It was proposed that such events could be identified through a double-peaked lightcurve created by the fallback of the two debris streams. However, this feature is unlikely to be observationally distinguishable because the time delay between the disruptions is generally small compared to the timespan of each individual TDE. It remains possible that the overall lightcurve displays a change of slope, but only if either the properties or the amount of mass loss differ significantly between the two stars \citep{mainetti2016}.

The above mechanisms produce tidal disruptions with a time delay between them that approaches the duration of a single TDE. This implies that the two events may not be completely independent. In this paper, we focus on such double TDEs and explore the possibility of collision between the two streams produced by each individual disruption \textit{before} they come back to the black hole. Staying agnostic about the mechanism at the origin of the double TDE, we analytically derive conditions on the stellar trajectories for streams collision to occur. If the two disruptions follow the tidal separation of a binary, we find that streams collision can happen with a probability of up to $44\%$. Using smoothed-particle-hydrodynamics (SPH) simulations, we confirm the validity of our analytical estimates and demonstrate that streams collision results in the formation of shocks that heat the gas. If radiation is able to promptly escape, this interaction could be detected as a burst of radiation with a luminosity $\sim 10^{43} \ergs$ lasting for at least a few days. We argue that this signal could act as a precursor of the main flare of TDEs and therefore be used to get a better handle on the different phases of these events such as the accretion disc formation process from observations.

The paper is organized as follows. Section \ref{collision} starts by presenting analytical conditions for streams collision to occur during a double TDE without specifying the mechanism that creates it. These conditions are then used in Section \ref{binary_likelihood} to compute the likelihood of this outcome for a specific mechanism involving the tidal separation of a stellar binary. In Section \ref{simulations}, hydrodynamical simulations of double TDEs are presented to test our analytical conditions and determine the impact of streams collision on the gas evolution. Finally, we discuss the results and present our conclusions in Section \ref{discussion}.

\begin{figure*}
\includegraphics[width=\textwidth]{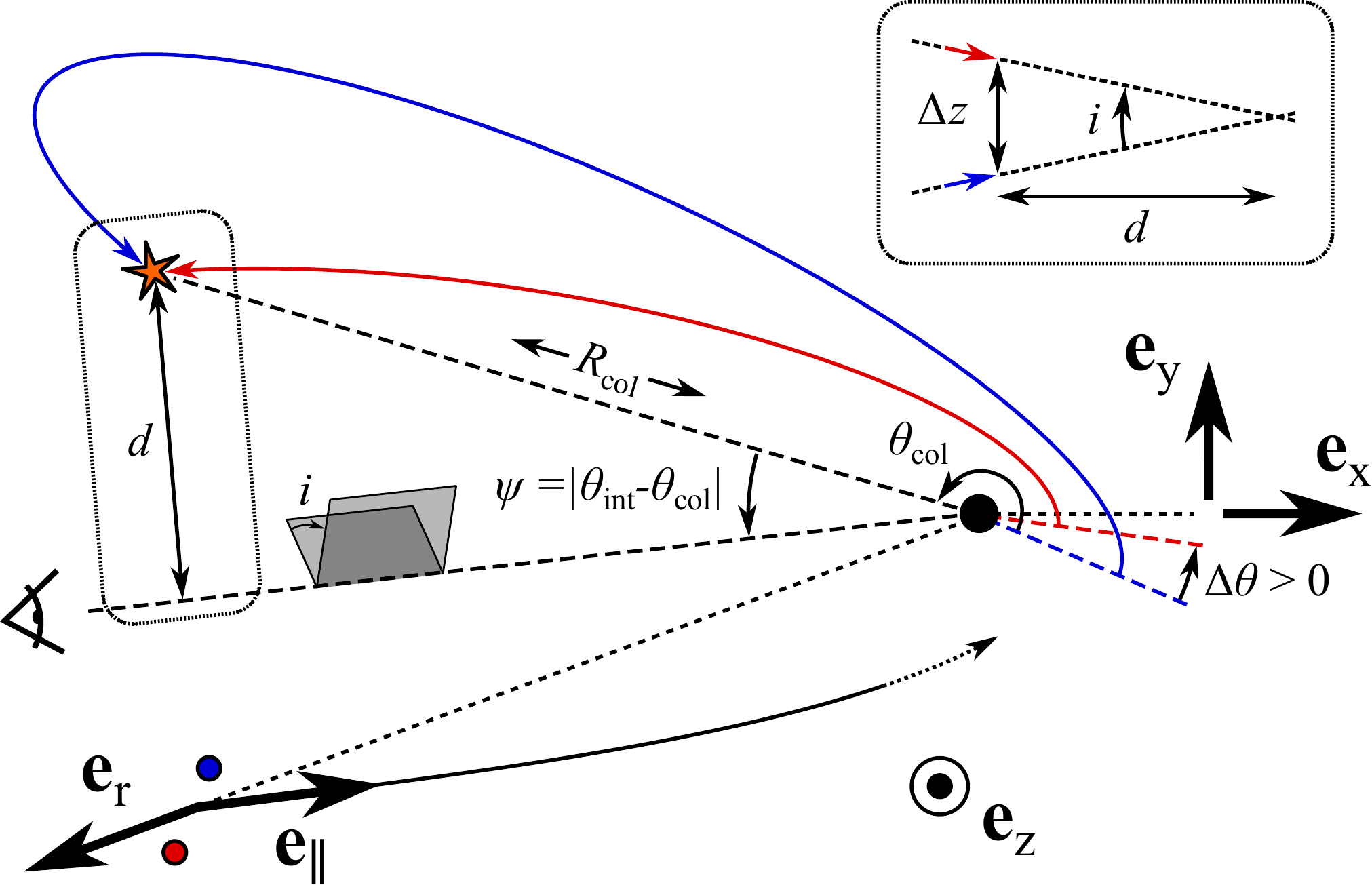}
\caption{Sketch illustrating the analytical treatment used to estimate the conditions for streams collision. The black hole position is marked by the big black dot. The first star to reach its pericenter is depicted by a blue point while the second star is represented by a red point. Their center of mass approaches the black hole along the black solid line following the dotted arrow. After the disruptions, the two streams revolve around the black hole. The blue and red lines show the trajectories of an element of the first and second stream, respectively. In the situation depicted here, the pericenter shift satisfies the condition $\dtheta\geq0$ for streams collision. If they evolve on the same plane, the two elements collide at the location of the orange star. This collision point is situated at a true anomaly $\thetacol$ and a distance $\rcol$ from the black hole. If the streams evolve on different orbital planes, these planes intersect along a line that passes through the black hole at a true anomaly $\thetaint$. This intersection line makes an angle $\psi = |\thetaint-\thetacol|$ with the direction connecting the black hole and the collision point. The collision point is located a distance $d$ from the intersection line, moving perpendicular to it. The inset in the upper right corner shows the trajectory of the two streams seen along the direction of the plane intersection line. The two orbital planes are inclined by an angle $i$ that determines the vertical offset $\Delta z$ at the collision point. The two elements collide only if $\Delta z<H$, where $H$ denotes the width of the streams.}
\label{sketch_collision}
\end{figure*}

\section{Streams collision}
\label{collision}

The sequential tidal disruption of two stars results in two debris streams that revolve around the black hole. These streams may interact with each other before they return to pericenter likely resulting in a modification of their dynamics and the emission of an electromagnetic signal that are specific to double TDEs. Remaining agnostic about the origin of the two disruptions, we start by analytically estimating the conditions for such a collision to happen (Section \ref{conditions}) and then determine its spatio-temporal evolution (Section \ref{evolution}) based on the incoming stellar trajectories.

For simplicity, we assume that the disrupted stars are identical, with the same mass $\mstar = \msun \ms$ and radius $\rstar = \rsun \rs$. They also follow parabolic orbits with the same direction of rotation. The stars are tidally disrupted if they reach a distance from the black hole smaller than their common tidal disruption radius
\be
\rtdis = \rstar \left(\frac{\mh}{\mstar} \right)^{1/3} =  0.47 \, \rs \m6^{1/3} \ms^{-1/3} \, \au,
\label{tidal_disruption_radius}
\ee
where $\mh = 10^6 \m6 \msun$ denotes the black hole mass.\footnote{The variable representing the tidal disruption radius has a superscript `dis' to differentiate it from the tidal separation radius defined in equation \eqref{tidal_separation_radius} and whose corresponding variable has a superscript `sep'.} The depth of each encounter is given by the penetration factor
\be
\beta = \rtdis/\rp \geq 1,
\ee
where $\rp$ denotes the pericenter distance. This factor can differ for the two disruptions, taking two distinct values $\beta_1$ and $\beta_2$. Upon each disruption, an energy spread 
\be
\Delta \epsilon = \frac{G \mh}{{\rtdis}^2}\rstar,
\label{energy_spread}
\ee
is imparted to the debris. As a result, the gas distributions evolve into elongated streams within which roughly half of the debris is unbound and escapes on hyperbolic orbits while the rest remains bound on elliptical orbits. The energy of the debris varies between $-\Delta \epsilon$ for the most bound one and $\Delta \epsilon$ for the most unbound. The former has a semi-major axis
\be
\amin = \frac{G \mh}{2 \Delta \epsilon} = 23 \, \m6^{2/3} m^{-2/3}_{\star} r_{\star} \au,
\label{semi_major_axis}
\ee
and returns to pericenter a time
\be
\tmin = 2 \pi \left(\frac{\amin^3}{G \mh}\right)^{1/2} = 41 \, M^{1/2}_6 m^{-1}_{\star} r^{3/2}_{\star} \d,
\label{fallback_time}
\ee
after the disruption of the star as predicted by Kepler's third law. The energy spread given by equation \eqref{energy_spread} does not depend on the penetration factor of the disruption because it is set at the tidal disruption radius \citep{sari2010,stone2013}. As a result, the range of energies of the debris is the same for the two streams, independently of the values of $\beta_1$ and $\beta_2$. A dependence on the penetration factor is however present in the eccentricity of the debris, given by
\be
e_{\rm min} = 1-\frac{2}{\beta}\left(\frac{\mh}{\mstar}\right)^{-1/3},
\label{most_bound_eccentricity}
\ee
\be
e_{\rm max} = 1+\frac{2}{\beta}\left(\frac{\mh}{\mstar}\right)^{-1/3},
\label{most_unbound_eccentricity}
\ee
for the most bound and most unbound, respectively. This means that these eccentricities differ between the two stars if they have different pericenter distances. Except when otherwise specified, we nevertheless assume that the penetration factors have the same value $\beta$ in this section.

\subsection{Conditions for collision}
\label{conditions}

We now derive analytical conditions for the two debris streams produced by the disruptions to collide with each other before they come back to the black hole. Here, we assume that the two stars have the same penetration factor $\beta =1$ and defer the complication associated with different pericenter distances to Appendix \ref{varying_beta} for clarity.

\subsubsection{Coplanar streams}

As a first step, we assume that the two streams evolve in the same orbital plane with aligned angular momentum vectors. In this case, the condition for streams collision can be expressed in terms of two quantities, which we refer to as $\dt$ and $\dtheta$. The first, $\dt$, is positive and denotes the time delay between the passages of the two stars at pericenter. In the remaining of the paper, we define the `first' star as the one that reaches pericentre first. Similarly, the star that passes the second at pericenter is referred to as the `second' star. The associated streams are named in the same way. Using these definitions, the time delay can be written as
\be
\dt = t_{\rm p,2} - t_{\rm p,1} \geq 0,
\ee
where $t_{\rm p,1}$ and $t_{\rm p,2}$ are the times of arrival at pericenter of the first and second star, respectively. The second quantity, $\dtheta$, can either be positive or negative. It is an angle that measures the relative shift in the pericenter location of the two stars. It is computed from
\be
\dtheta = \theta_2 - \theta_1,
\label{dtheta}
\ee
where the angles $\theta_1$ and $\theta_2$ measure the pericenter location of the first and second star, respectively, with respect to a reference direction. As a convention, we impose these angles to increase in the common direction of rotation of the stars. The angle given by equation \eqref{dtheta} is also the pericenter shift between any element of the first stream and any element of the second stream since the debris follows ballistic orbits.%

\begin{figure*}
\includegraphics[width=0.47\textwidth]{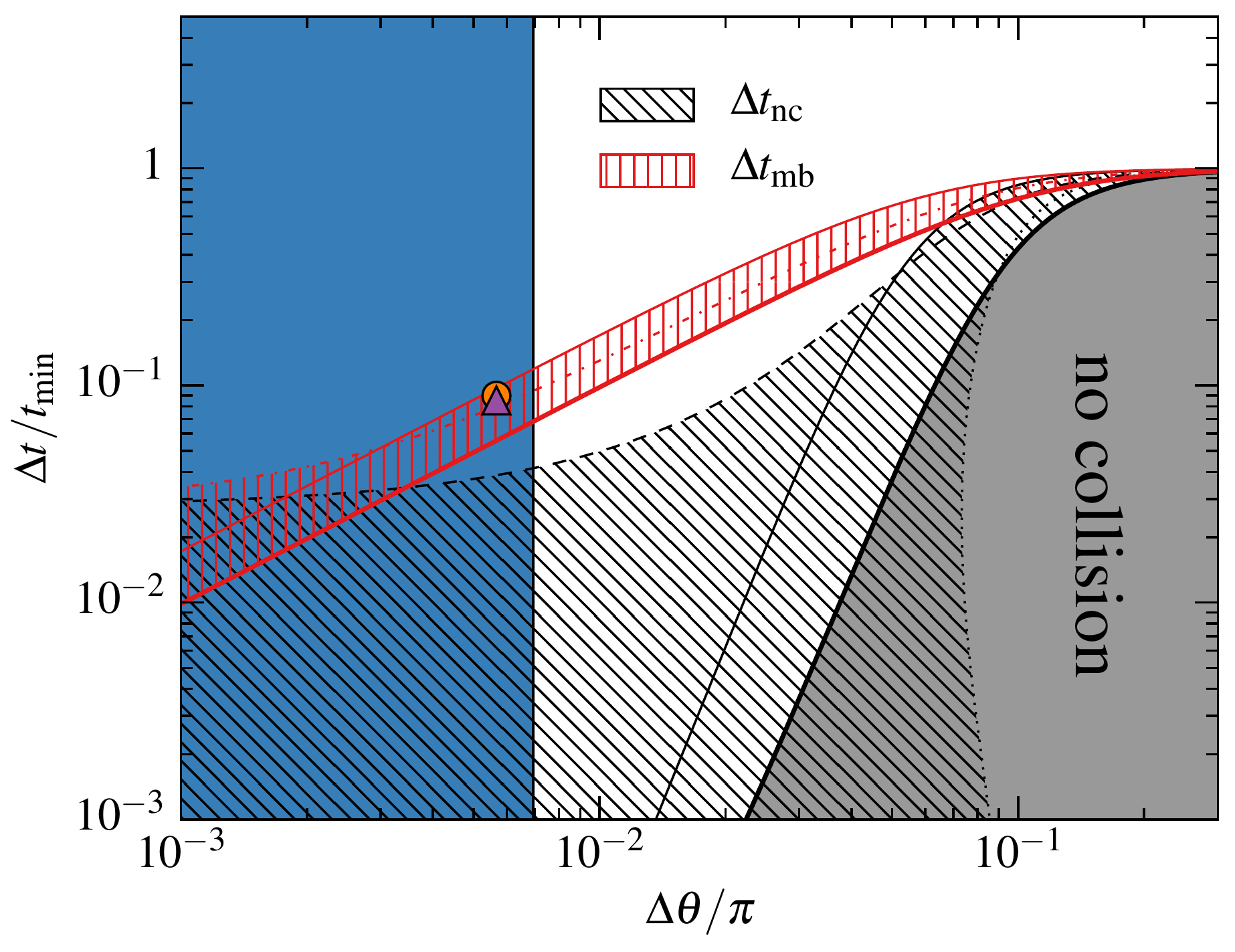}
\hspace{0.02\textwidth}
\includegraphics[width=0.47\textwidth]{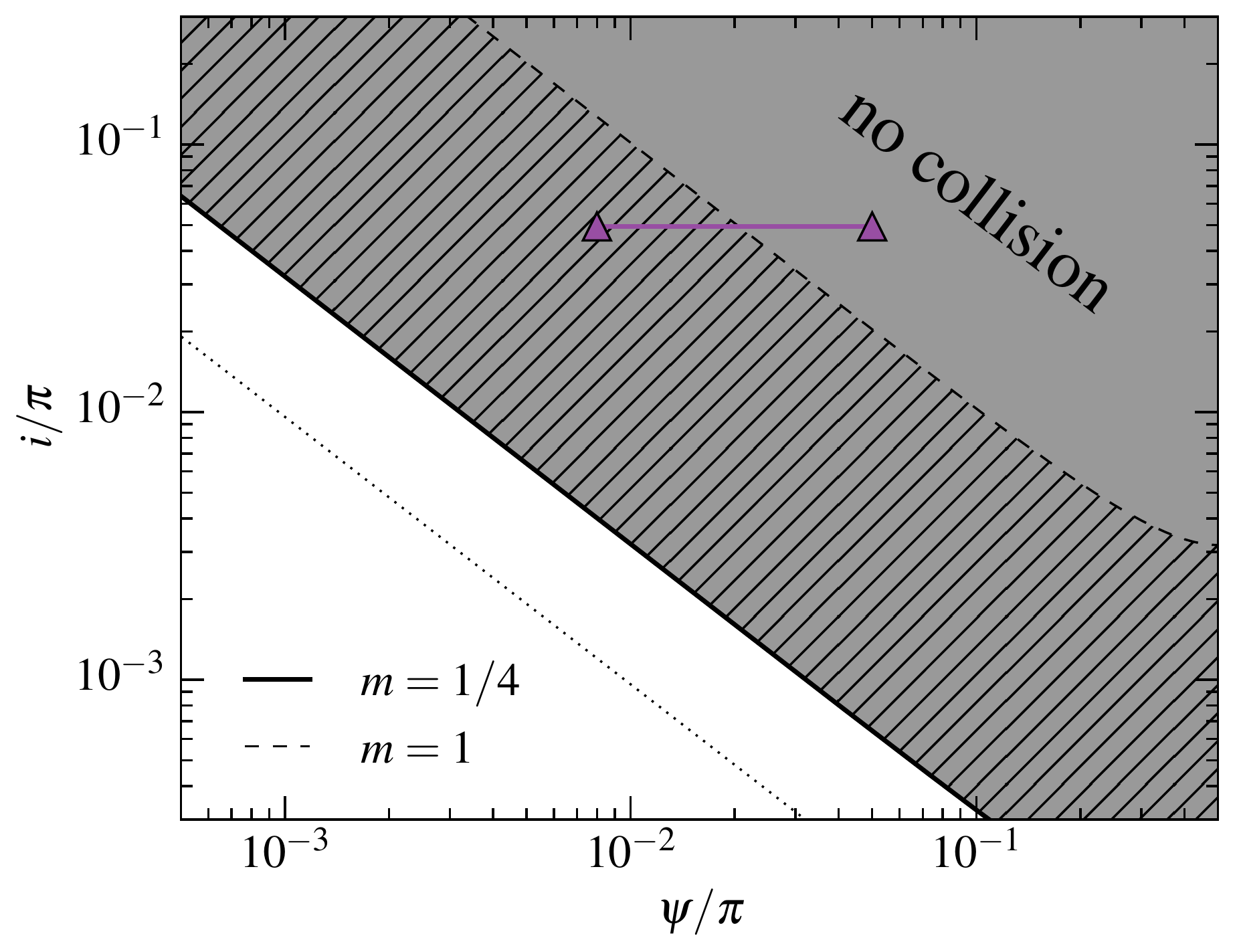}
\caption{Conditions for streams collision shown in the $\dtheta$-$\dt$ (left panel) and $\psi$-$i$ (right panel) planes. If the two streams evolve on the same plane with $\dtheta \geq0$, collision can still be avoided if the second stream is able to pass between the most bound debris of the first stream and the black hole. For $\beta_1=\beta_2=1$, this happens inside the grey area of the left panel delimited by the black thick solid line while the red thick solid line corresponds to a collision between the most bound debris of each stream. For penetration factors both varying between 1 and 3, the black and red lines move inside the hatched regions of the same colour. The characteristics of each line interior to these areas are explained in Appendix \ref{varying_beta}. If the streams evolve on different orbital planes, they can pass on top of each other instead of colliding. For streams confined by self-gravity ($m=1/4$), collision is avoided in this way inside the grey area of the right panel delimited by the black thick solid line if the collision radius is $\rcol = 2 \, \amin$. The boundary of this region moves downward to the dotted black line if the collision radius is $\rcol = 10 \, \amin$ and upward to the dashed black line if the streams expand homologously ($m=1$). The blue region in the left panel is covered for two disruptions resulting from a previous binary separation in the Keplerian case. The initial conditions of our simulations are indicated in the two planes by the orange circle (model R$\alpha$0) and purple triangles (model R$\alpha$0.1).}
\label{planes}
\end{figure*}

A first condition for the two streams to interact is $\dtheta \geq 0$. This means that the pericenter location of the second star is further in the direction of motion than that of the first star. Equivalently, the second stream has a major axis more rotated than that of the first one in this forward direction. As a result, the second stream can catch up with the first stream resulting in a collision between a fraction of their elements before they come back to pericenter.\footnote{In fact, streams collision can also happen for $\dtheta<0$ in some special circumstances. One possibility is that the second star passes closer to the black hole than the first. In this case, the shorter time spent close to pericenter by the second stream can cause it to get ahead of the first stream even if the second star was initially delayed. This type of collisions is however less likely than the ones with $\dtheta\geq0$ and not considered in the following of the paper.} This is illustrated in Fig. \ref{sketch_collision} where the trajectories of elements of the first and second stream are shown with blue and red arrows, respectively. Since $\dtheta \geq 0$, these two trajectories cross at the location of the orange star. Additionally, it takes longer for the first element to reach that position than for the second. These two elements are therefore able to reach the collision point at the same time owing to $\dt\geq0$.

Assuming that the inequality $\dtheta \geq 0$ is verified, streams collision can still be avoided if the pericenter shift $\dtheta$ is so large that the second stream is able to pass between the black hole and the most bound debris of the first stream.\footnote{In this case, a collision can still happen very close to pericenter when both streams come back to the black hole vicinity. We do not consider this type of collisions in the rest of the paper since it is unlikely to change the overall dynamics of the streams. However, it could affect the formation of an accretion disc from this gas.} This is only possible if this most bound element has not yet fallen back to pericenter when the second star is disrupted, that is if $\dt \leq \tmin$. The borderline case is a situation where the most bound debris of the first stream interacts with the most unbound debris of the second. This corresponds to a region of the $\dtheta$-$\dt$ plane delimited by a function parametrized by $\thetacol$, which denotes the true anomaly at which the collision happens, measured from the pericenter location of the first star (see Fig. \ref{sketch_collision}) and varying between 0 anf $2 \pi$. The associated pericenter shift $\dthetanc$ is obtained by imposing that the most unbound element of the second stream reaches the same radial position as the most bound element of the first stream. This condition can be written $(1+e_{\rm min})/(1+e_{\rm min} \cos \thetacol) = (1+e_{\rm max})/(1+e_{\rm max} \cos (\thetacol-\dthetanc))$, which uses the assumption of equal penetration factors for the two stars and the fact that the collision happens at a true anomaly $\thetacol$ for the first stream and $\thetacol - \dthetanc$ for the second. The solution of the above condition is
\be
\dthetanc = \thetacol - \arccos \bigg\{\frac{1}{e_{\rm max}} \left[\frac{(1+e_{\rm max})(1+e_{\rm min} \cos \thetacol)}{1+e_{\rm min}} -1\right]\bigg\}.
\label{dthetanc}
\ee
For pericenter shifts larger than this critical value, there is no collision. The solution can be written as in equation \eqref{dthetanc} because the true anomaly of the element of the second stream satisfies the condition $\thetacol-\dthetanc<\pi$ due to the fact that it is unbound and only moves outwards. The corresponding time delay is computed by imposing that the two stream elements reach the collision point at the same time, which gives
\be
\dtnc = t_1(-\Delta \epsilon,\thetacol) - t_2(\Delta \epsilon,\thetacol - \dthetanc),
\label{dtnc}
\ee
where $t_1 (\varepsilon,\theta)$ and $t_2 (\varepsilon,\theta)$ represent the time needed for a gas element of energy $\epsilon$ to reach a position on its orbit corresponding to a true anomaly $\theta$ if it belongs to the first and second stream, respectively. The parametric function defined by equations \eqref{dthetanc} and \eqref{dtnc} traces a line in the $\dtheta$-$\dt$ plane. It is represented by the thick solid black curve that delimits the grey area in the left panel of Fig. \ref{planes}. Inside this region, the streams do not collide.

Outside the grey region, a collision takes place between the two streams. Its outcome then depends on the location in the $\dtheta$-$\dt$ plane. To understand how, it is first instructive to examine the situation where the most bound debris of the two streams collide with each other. It corresponds to a line in the plane given by a parametric function that can be derived in the same way as above. The condition for the two elements to reach the same radial position at the collision point reduces to $\cos \thetacol = \cos (\thetacol-\dthetamb)$ as obtained by replacing $e_{\rm max}$ by $e_{\rm min}$ in equation \eqref{dthetanc}. As before, $\thetacol$ denotes the true anomaly of the first stream element at the collision point. The solution of this equality is
\be
\dthetamb = 2(\thetacol-\pi),
\label{dthetamb}
\ee
that is the value of the pericenter shift for which the most bound part of each stream collide together. The solution takes this form because the collision occurs while the element of the first stream moves inwards with $\thetacol>\pi$ and that of the second streams moves outwards with $\thetacol-\dthetamb>\pi$.
The associated time delay is
\be
\dtmb = t_1(-\Delta \epsilon,\thetacol) - t_2(-\Delta \epsilon,\thetacol-\dthetamb),
\label{dtmb}
\ee
which is obtained by imposing that the two most bound elements reach the collision point at the same time. The function defined by equations \eqref{dthetamb} and \eqref{dtmb} is shown with a thick solid red line in the left panel of Fig. \ref{planes}. Along this curve, a collision takes place between the most bound debris of the streams.

It is now possible to estimate the outcome of the streams collision as a function of $\dt$ and $\dtheta$. One particularly important characteristic is the collision strength that depends on the fraction of streams involved and their relative speed when they collide. It varies with the position in the $\dtheta$-$\dt$ plane of Fig. \ref{planes} with respect to the red line determined above. Along this line, the two most bound elements collide with each other implying that the bound fraction of streams involved in collision is maximized at fixed $\dtheta$. Above it, part of the the first stream bound debris avoids the collision because, owing to the larger $\dt$, it has already passed the collision region before the second stream arrives. Below the line, the opposite happens and some of the second stream bound elements do not interact. Depending on the location on that line, different regimes of collision also exist. On the right-hand side, where $\dt \approx \tmin$ and $\dtheta \approx 0.3 \pi$, the collision occurs between a still compact second stream that passes through a tenuous and extended first stream. The interaction therefore happens at high velocity but remains weak due to the large density ratio. On the left-hand side, where $\dt \ll \tmin$ and $\dtheta \ll \pi$, the two streams are moving on very similar trajectories. The collision therefore involves a large fraction of the two streams, but the relative velocity is small. This qualitative analysis suggests that the strongest collisions are to be expected close to the red line and in between these two regimes.

\subsubsection{Effect of inclination}

We now treat the more general case where the two streams do not evolve on the same orbital plane. In this situation, it is possible that they pass on top of each other instead of colliding. To estimate the condition for interaction, we compare the vertical offset induced by the orbital plane inclination to the width of the streams at the collision point. Note that we keep referring to this location at the `collision point' even though the collision may not happen due to the vertical offset between the streams. The different variables used to perform this estimate are shown in Fig. \ref{sketch_collision}. For orbital planes inclined by a small angle $i$, the vertical offset is given by $\Delta z = d i$ where $d$ is the distance of the collision point to the intersection line moving perpendicular to it. Estimating this distance requires to know the position of the collision point with respect to the intersection line of the two planes. The collision happens at a true anomaly $\thetacol$. In addition, we define the true anomaly $\thetaint$ of the plane intersection line, which is possible since it passes through the black hole. As for $\thetacol$, this true anomaly is measured from the location of the first star pericenter and increases in the direction of motion.\footnote{Since the orbital planes of the streams are different, the true anomalies $\thetacol$ and $\thetaint$ can be measured on either planes. However, these two choices give essentially the same value because the planes are inclined by a small angle.} Using these definitions yields $d = \rcol \sin \psi$ where
\be
\psi = |\thetaint-\thetacol|,
\ee
is the positive angle that the plane intersection line makes with the direction connecting the collision point and the black hole. It is then possible to compute the vertical offset as
\be
\Delta z = \rcol \, i \sin \psi.
\ee
The next step is to evaluate the width of the streams at the collision point. This width can be estimated as
\be
H = \rstar \left(\frac{\rcol}{\rtdis}\right)^m,
\ee
where $m$ depends on which mechanism sets it. If the width is determined by hydrostatic equilibrium between gas pressure and self-gravity, the slope is $m=1/4$. If it is instead set by tidal forces, the evolution is homologous with $m=1$ \citep{kochanek1994,coughlin2016-structure}. While hydrostatic equilibrium is maintained in most of the stream for weak encounters with $\beta \approx 1$, an homologous evolution is expected if thermal energy is injected into the gas during the disruption, which requires $\beta \gtrsim 3$. The ratio of vertical offset to streams width is then
\be
\begin{split}
\frac{\Delta z}{H}
& = \left(\frac{\mh}{\mstar} \right)^{(2-m)/3} \left(\frac{\rcol}{2 \, \amin} \right)^{1-m} i \, \sin \psi \\
& =
\begin{cases}
\m6^{-1/3} \ms^{1/3} \,\, \frac{i \sin \psi}{0.001 \pi^2}, & m=1\\
\m6^{-7/12} \ms^{7/12} \,\, \left(\frac{\rcol}{2 \, \amin}\right)^{3/4} \frac{i \sin \psi}{3 \times 10^{-5} \pi^2}. & m=1/4
\end{cases}
\end{split}
\label{condition_inclination}
\ee
For $m =1$, the collision radius $\rcol$ cancels out. For $m =1/4$, the numerical estimate assumes that the collision happens close to the apocenter of the streams most bound debris, that is $\rcol \approx 2 \, \amin$ using equation \eqref{semi_major_axis}. The borderline case $\Delta z=H$ is represented in the $\psi$-$i$ plane shown in the right panel of Fig. \ref{planes} by the black thick solid line for $m=1/4$ and assuming $\rcol=2\,\amin$. It delimits the grey area inside which the collision is avoided if the streams expansion is confined by self-gravity. If $\rcol =10 \, \amin$, this area extends downwards to the black dotted line. The boundary of that region is indicated by the dashed black line if the stream expands homologously with $m=1$. The hatched area denotes its location for an intermediate case with $1/4<m<1$ that is obtained for example when one stream evolves homologously while the other one is confined by self-gravity. As expected, the streams are more likely to interact if their width evolves homologously than if it is confined by self-gravity. Nevertheless, this interaction is weakened by the vertical offset between the streams because it prevents part of the gas from colliding.

In summary, we have derived three conditions for streams produced by double disruptions to collide with each other. The first condition requires that the pericenter shift $\dtheta$ between the stars is positive in order for the second stream to be able to catch up with the first one. The second condition imposes that this shift is lower than the critical value $\dthetanc$ if the time delay is smaller than $\tmin$ to prevent the second stream from entirely passing between the first stream and the black hole. The third condition applies if the two streams evolve on inclined orbital planes, in which case the vertical offset induced by this inclination must be smaller than the streams width for a collision to happen.

\subsection{Evolution of the collision point}

\label{evolution}

One can also get insight into the spatio-temporal evolution of the collision point that is continuously reached by different parts of each stream. This analysis assumes that the most bound debris of each stream collide with each other, which corresponds to the red line in the $\dtheta$-$\dt$ plane of Fig. \ref{planes}. Away from this line, it nevertheless provides a good estimate for the location of the collision point. Typically, the first collision happens between the most bound elements while the last one involves the most unbound element of the second stream. The true anomaly at the location of these collisions can be estimated as 
\be
\thetacol^{\rm mb} \approx \pi+\dtheta/2,
\label{thetacolmb}
\ee
\be
\thetacol^{\rm mu} \approx \pi +\dtheta - \gammamu,
\label{thetacolmu}
\ee
for the first and last one, respectively. The first angle is obtained by reverting equation \eqref{dthetamb}. The second angle uses the fact that the second stream most unbound element reaches the collision point while its trajectory is already approximately straight. This trajectory is inclined with respect to the major axis of the second stream by a angle $\gammamu = \arccos(1/e_{\rm max})$ given by
\be
\gammamu \approx 2 \, \beta^{-1/2} \left(\frac{\mh}{\mstar}\right)^{-1/6} \approx 0.064 \, \pi \, \beta^{-1/2} \m6^{-1/6} \ms^{1/6},
\label{gammamu}
\ee
making use of the small angle approximation and equation \eqref{most_unbound_eccentricity}. As long as $\dtheta<2\gammamu \approx 0.13 \pi$, the true anomaly of the collision point is therefore constrained to the interval
\be
\thetacol^{\rm mu} \leq \thetacol \leq \thetacol^{\rm mb}.
\label{thetacol_range}
\ee
For a pericenter shift $\dtheta \ll \pi$, equation \eqref{thetacolmb} implies that the most bound elements collide with each other a time
\be
\frac{\tcol^{\rm mb}}{\tmin} \approx 0.5,
\label{minimal_collision_time}
\ee
after the disruption of the first star and at a distance from the black hole $\rcol^{\rm mb} \approx 2 \, \amin$, that is close to the apocenter of the streams most bound debris. The associated emission could therefore act as a \textit{precursor} of the main flare accompanying the gas fallback at pericenter.

The most unbound debris of the second stream collides when it reaches an element of the first stream. As long as $\dtheta \leq \gammamu$, this element is also unbound but escapes with a slower velocity that allows the second stream to catch up with it. The collision happens after a time $t_{\rm col}^{\rm mu}$ given by the condition $\Delta v (t_{\rm col}^{\rm mu}-\dt) = v^{\rm col}_1 t_{\rm col}^{\rm mu}$ imposing that the two elements reach the same radial position. Here, $\Delta v$ and $v^{\rm col}_1$ denote the velocity of the second stream most bound debris and the colliding element of the first stream, respectively. Approximating these velocities by their value at infinity, they are related to their respective energies by $v^{\rm col}_1/\Delta v \approx (\epsilon^{\rm col}_1/\Delta \epsilon)^{1/2}$. This energy ratio can be computed from the fact that the two colliding elements must follow the same straight line as they escape from the black hole. Denoting by $e^{\rm col}_1$ the eccentricity of the first stream element, this condition translates into $\arccos(1/e^{\rm col}_1)=\arccos(1/\emax_2)-\dtheta$ that gives $1-(\epsilon^{\rm col}_1/\Delta \epsilon)^{1/2} \approx (1/2) \beta^{1/2} \Delta \theta (\mh/\mstar)^{1/6}$ using the small angle approximation. The time at which the most unbound element of the second stream collides is therefore
\be
\begin{split}
\frac{\tcol^{\rm mu}}{\tmin}
& = \frac{\dt}{\tmin} \left( 1- \frac{v^{\rm col}_1}{\Delta v} \right)^{-1}  \\
& \approx \frac{2}{\dtheta \beta^{1/2}} \frac{\dt}{\tmin} \left( \frac{\mh}{\mstar} \right)^{-1/6} \\
& \approx 6.4 \, \beta^{-1/2} \m6^{-1/6} \ms^{1/6} \left( \frac{\dtheta}{10^{-2}\pi} \right)^{-1} \frac{\dt}{\tmin},
\end{split}
\label{maximal_collision_time}
\ee
which corresponds to a distance from the black hole of $\rcol^{\rm mu} \approx \Delta v \, t_{\rm col}^{\rm mu} = 40  \, \amin \, \beta^{-1/2} \m6^{-1/6} \ms^{1/6} (\dtheta/10^{-2} \pi)^{-1} \dt/\tmin$. This calculation demonstrates that streams collision can be sustained for a long duration and happen far away from the black hole.

\begin{figure}
\centering
\includegraphics[width=0.9\columnwidth]{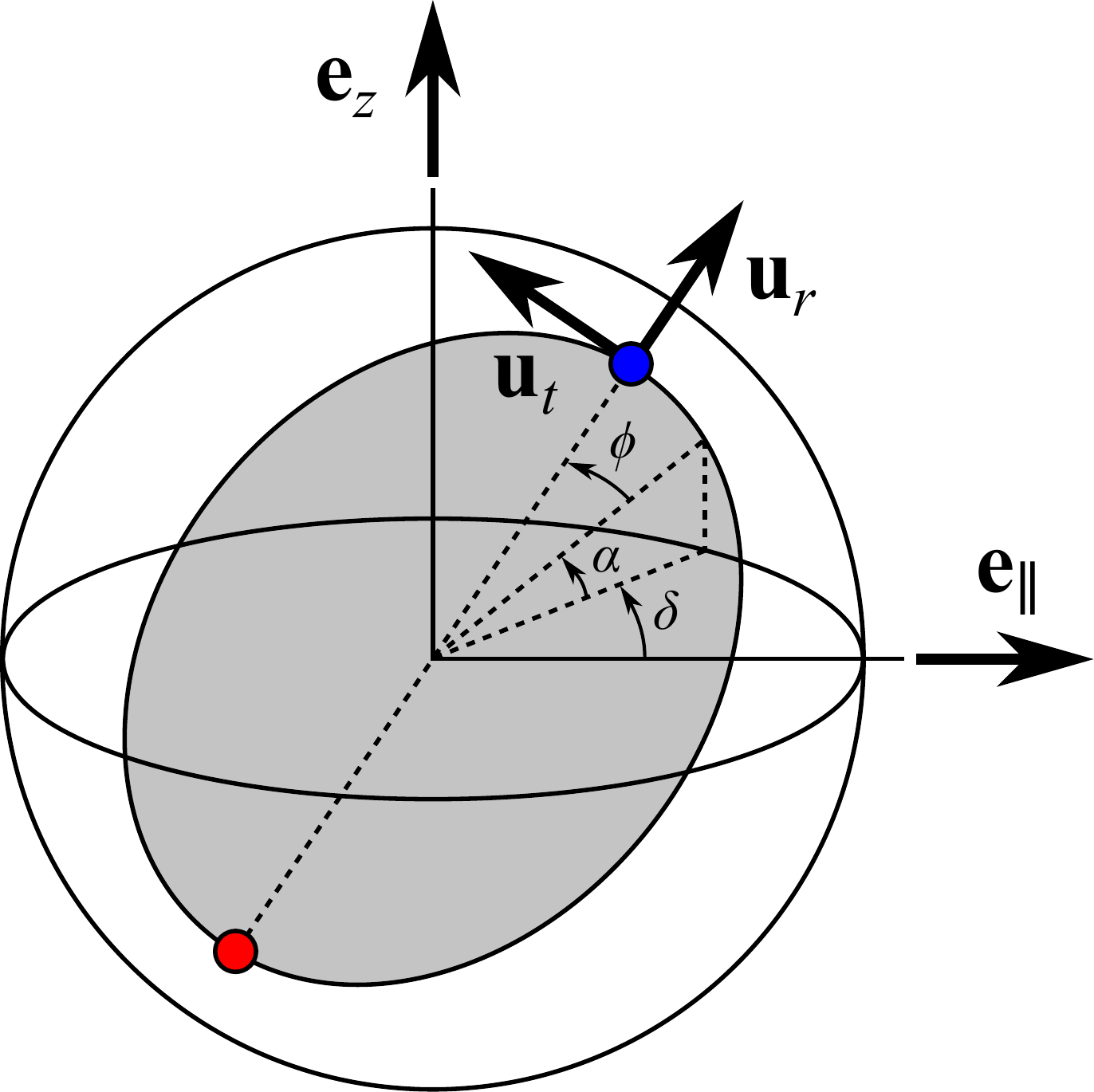}
\vspace{0.2cm}
\caption{Configuration of the binary at the moment of separation. The first star is represented by a blue point and the second star by a red point. The pitch angle $\alpha$ and yaw angle $\delta$ fix the orbital plane of the two stars while the phase angle $\phi$ defines the position of the first star in this plane. The vector $\epara$ indicates the direction of motion of the binary center of mass.}
\label{sketch_binary}
\end{figure}

\section{Binary tidal separation}
\label{binary_likelihood}

We now want to evaluate the likelihood of streams collision in the particular case where the two disruptions are due to the tidal separation of a stellar binary. This is done by first evaluating the values of $\dt$, $\dtheta$, $i$ and $\psi$ and their dependence on the binary properties. The likelihood of streams collision is then obtained by using the conditions derived in the previous section and summarized in Fig. \ref{planes}. 

The two identical stars considered in Section \ref{collision} are now part of a binary that we assume for simplicity to be circular. These two binary components are separated by the tidal force of the black hole at the tidal separation radius
\be
\rtsep = a \left(\frac{\mh}{2 \,\mstar} \right)^{1/3} = 370 \, \a3  \m6^{1/3} \ms^{-1/3} \au,
\label{tidal_separation_radius}
\ee
where $a= 10^3 \a3 \rsun$ is the binary separation. It is possible to estimate the range of values that this separation can take. A lower limit is given by $a \gtrsim 2 \rstar$ to prevent the two stars from colliding with each other. An upper limit is set by ionization of the binary through two-body encounters with surrounding stars. To estimate this limit, we adopt the properties of the Milky Way nuclear star cluster described by \citet{antonini2011} in their section 2. In particular, we use the same density profile and velocity dispersion. The critical radius from  which most tidally separated binaries originate can be evaluated using the loss cone theory, described for instance by \citet{syer1999} (section 3) in the context of stellar tidal disruptions. We find this radius to be $R_{\rm crit} \gtrsim \kpc$ for separations $a \gtrsim 10^3 \rsun$, which is much larger than for single tidally disrupted stars. From equation (7.173) of \citet{binney2008}, the ionization timescale $t_{\rm ion}$ at that distance can be shown to be longer than a stellar lifetime, meaning that binaries survive ionization. Another possibility is that the binary is ionized while it is already on a highly eccentric orbit grazing the tidal separation radius. In this situation, the upper limit on the binary separation is reached when the integral $\int_{R_{\rm crit}}^{\rtsep} \diff t/t_{\rm ion}$ increases to a value close to 1. This condition imposes that the binary is ionized after a few near-radial oscillations between the critical radius and the tidal separation radius. It yields an upper limit of $a \lesssim 10^5 \rsun$ for the binary separation. In the remaining of this section, we adopt $a=10^3 \rsun$ in numerical estimates as a typical binary separation value.

The binary center of mass reaches the tidal separation radius given by equation \eqref{tidal_separation_radius} following a parabolic orbit with a pericentre $\rp = \rtdis/\beta$, where $\beta \geq 1$ is the penetration factor of the binary with respect to the tidal disruption radius of its components (equation \ref{tidal_disruption_radius}). The individual penetration factors of the stars $\beta_1$ and $\beta_2$ are set by the change in angular momentum imparted by the separation. This variation of angular momentum becomes similar to that of the center of mass only for wide binaries with $a/\rstar \gtrsim \beta^{-1} (\mh/\mstar)^{2/3} \approx 10^4 \beta^{-1} \m6^{2/3} \ms^{-2/3}$, for which the orbit of one of the two stars can flip. Otherwise, the stars retain the same direction of rotation with similar penetration factors. The energy of the stars also get modified by the separation process.  Nevertheless, this variation is smaller than the energy spread $\Delta \epsilon$ induced by the tidal disruptions (equation \ref{energy_spread}) by a factor $\sim a/\rstar = 10^3 \a3 \rs^{-1}$. It therefore has a negligible influence on the dynamics of the debris streams and we can consider the two stars to be disrupted on parabolic orbits as assumed in Section \ref{conditions}.

\subsection{Keplerian case}

Assuming that the two stars follow perfectly Keplerian orbits, the quantities involved in the conditions for streams collision are entirely determined by the change in trajectory experienced by the stars during binary separation. This process happens on a time-span very short compared to the binary period. It can therefore be approximated as an instantaneous deflection of the two stars on new trajectories happening at the tidal separation radius. The position and velocity vectors of the binary with respect to the black hole at the moment of separation are given by 
\be
\vect{R} = \rtsep \er,
\label{position_com}
\ee
\be
\vect{v} = \left( \frac{2G\mh}{\rtsep} \right)^{1/2} \epara,
\label{velocity_com}
\ee
respectively. Here, $\er$ and $\epara$ are unit vectors shown in Fig. \ref{sketch_collision} that indicate the radial direction of the binary center of mass and its direction of motion, respectively. The change in trajectory suffered by the first star is dictated by the displacement and velocity kick
\be
\dr = \frac{a}{2} \, \ur,
\label{position_kick}
\ee
\be
\dv = \left(\frac{G \mstar}{2 \, a} \right)^{1/2} \ut,
\label{velocity_kick}
\ee
with respect to the center of mass trajectory. They are simply given by the position and velocity of this star with respect to the binary center of mass at the moment of separation. The second star experiences a displacement and velocity kick of the same magnitude but opposite direction, given by $-\dr$ and $-\dv$. As depicted in Fig. \ref{sketch_binary}, the directions of the unit vectors $\ur$ and $\ut$ are determined by three random angles $\alpha$, $\delta$ and $\phi$. The pitch angle $\alpha$ and yaw angle $\delta$ set the orientation of the plane in which the two stars rotate while the phase angle $\phi$ fixes the position of the first star in that plane. For example, $\alpha =0$ corresponds to a situation where the two stars move on a plane that is identical to that of the binary center of mass around the black hole. These two planes are instead perpendicular for $\alpha = \pi/2$. The fact that $|\dr|/|\vect{R}| \approx |\dv|/|\vect{v}| \approx (\mh/\mstar)^{-1/3} \ll 1$ implies that the change of trajectory induced by the separation process is small. This variation can therefore legitimately be computed at first order in the displacement and velocity kick, which we will do in the following.

Using these definitions, we first derive the time delay $\dt$ given by the separation process. To reach pericenter first, the first star must be closer to the black hole than the second one at the moment of separation. The associated time delay can be evaluated as $\dt \approx a \eta /|\vect{v}|$, where $\eta =\ur \cdot \epara$ corresponds to the projection of the separation on the binary orbital plane. The ratio of this delay to the fallback time given by equation \eqref{fallback_time} is then
\be
\begin{split}
\frac{\dt}{\tmin} & \approx \frac{1}{2^{1/6} \, \pi} \, \eta \left( \frac{a}{\rstar} \right)^{3/2} \left( \frac{\mh}{\mstar} \right)^{-5/6} \\
& = 0.09 \,\, \eta \a3^{3/2} \rs^{-3/2} \m6^{-5/6} \ms^{5/6},
\end{split}
\label{time_delay}
\ee
where $\eta = \cos \alpha \cos \delta \cos \phi - \sin \delta \sin \phi$ satisfies $0 \leq \eta \leq 1$. Note that, as expected, $\eta$ is independent of the direction of rotation of the two stars around each other. The upper limit $a \lesssim 10^5 \rsun$ on the binary separation leads to $\dt \lesssim 90 \, \tmin$. Note that the lower limit is irrelevant since $\eta$ can reach 0. This implies that the whole vertical extent of the $\dtheta$-$\dt$ plane is covered, as indicated with the blue region shown in the left panel of Fig. \ref{planes}.

Evaluating the rest of the parameters requires to know the variation in angular momentum and eccentricity vectors undergone by the binary components upon separation. The binary center of mass angular momentum and eccentricity vectors are
\be
\vect{j} = \vect{R} \times \vect{v} = \left( 2 G \mh \rp \right)^{1/2} \ez,
\ee
\be
\vect{e} = \frac{\vect{v} \times \vect{j}}{G \mh} - \frac{\vect{R}}{|\vect{R}|} = \ex,
\ee
respectively, where $\ex$ and $\ez$ are unit vectors shown in Fig. \ref{sketch_collision} that form a orthogonal basis with the unit vector $\ey$. The small displacement and velocity kick given by equations \eqref{position_kick} and \eqref{velocity_kick} translate into variations in the angular momentum and eccentricity vectors of the first star during the separation process. To first order, they are given by 
\be
\ddj \approx \dr \times \vect{v} + \vect{R} \times \dv,
\ee
\be
\de \approx \frac{1}{G \mh} \left( \dv \times \vect{j} +  \vect{v} \times \ddj \right) - \frac{\dr}{|\vect{R}|}.
\ee
The second star undergoes kicks $-\ddj$ and $-\de$ as required to conserve the angular momentum of the system.

We now describe how to obtain $\dtheta$, $i$ and $\psi$. These quantities are computed at lowest order in $\rp/\rtsep = 10^{-3} \beta^{-1} \a3^{-1} \rs$, which is valid except for very compact binaries and allows us to clearly reveal the dependence on the physical parameters of the problem. Physically, this simplification results from the fact that the parabolic trajectory of the binary is a straight line far from pericenter. This corresponds to the lowest order approximation in $\rp/\rtsep$ while higher order terms would take into account the curvature of the trajectory. The pericenter shift $\dtheta$ results from the change in eccentricity vector experienced by the stars. It is given by the angle between the eccentricity vector of the first star $\vect{e}+\de$ and that of the second one $\vect{e}-\de$, considering only the components of $\de$ along the orbital plane of the binary around the black hole. To first order, this results in
\be
\begin{split}
\dtheta & \approx 2 \left( \de \times \vect{e} \right) \cdot \ez \\
& \approx 2^{1/3} \, \xi  \left( \frac{\mh}{\mstar} \right)^{-1/3} \\
& = 0.0069 \, \pi \, (\xi/\sqrt{3}) \m6^{-1/3} \ms^{1/3},
\end{split}
\label{pericenter_shift}
\ee
where $\xi = \cos \delta (\sin \phi \mp \sqrt{2} \cos \phi) + \cos \alpha \sin \delta ( \cos \phi \pm \sqrt{2} \sin \phi)$ obeys $-\sqrt{3} \leq \xi \leq \sqrt{3}$. Here and in the remaining of this section, the upper signs correspond to a binary rotating in the prograde direction compared to its motion around the black hole while the lower signs correspond to the retrograde case.\footnote{Going from the prograde to the retrograde case is equivalent to making the substitutions $\alpha \rightarrow \pi -\alpha$, $\delta \rightarrow \delta+\pi$ and $\phi \rightarrow -\phi$.} If the second star has its pericenter further in the direction of motion than the first star, $\de$ is directed approximately along $-\ey$ and the first equality of equation \eqref{pericenter_shift} gives $\dtheta >0$ as expected. The pericenter shift obeys $\dtheta/\pi \lesssim 0.0069$, implying that it is limited to the leftmost region of the $\dtheta$-$\dt$ plane, shown in blue in the left panel of Fig. \ref{planes}. This angle is small enough to stay outside the grey region of the plane, except for near-contact binaries with $\dt \lesssim 10^{-5} \tmin$. Streams collision is therefore unlikely to be avoided in this way. $\dtheta$ is largely independent of $\alpha$ because the first term of $\xi$ dominates as long as $\tan \delta \cos \alpha \lesssim 1$. One can therefore study the sign of $\dtheta$ for $\alpha=0$ without loss of generality. The condition $\dtheta \geq 0$ then translates into $\phi+\delta \geq \pm\arctan(\sqrt{2}) \approx \pm 0.3 \pi$. For the first star to be closer to the black hole than the second, this effective phase angle must additionally belong to the interval $-\pi/2\leq\phi+\delta\leq\pi/2$. The condition of positive pericenter shift is realized for less and for more than half of this allowed interval in the prograde and retrograde case, respectively. Taking into account the random distribution of the phase angle, $\dtheta \geq 0$ is satisfied with a probability of $0.5\pm\arctan(\sqrt{2})/\pi$. However, since the binary is as likely to be prograde as it is to be retrograde, the two contributions cancel out such that the overall probability of this condition reduces to exactly $50\%$.

The orbital plane inclination $i$ is simply the angle between the angular momentum vector $\vect{j}+\ddj$ of the first star and that $\vect{j}-\ddj$ of the second. It is therefore given by
\be
\begin{split}
i & \approx  2\frac{|\ddj \times \vect{j}|}{\vect{j}^2} \\
& = 2^{-5/6} \chi |\sin \alpha| \, \beta^{1/2}  \left( \frac{a}{\rstar} \right)^{1/2} \left( \frac{\mh}{\mstar} \right)^{-1/3}\\
& = 0.14 \, \pi \, (\chi/\sqrt{6}) |\sin \alpha|  \, \beta^{1/2}  \a3^{1/2} \rs^{-1/2} \m6^{-1/3} \ms^{1/3},
\end{split}
\label{inclination}
\ee
where $\chi =  \sqrt{3 + \cos (2 \phi) \pm 2 \sqrt{2} \sin (2 \phi)} \leq \sqrt{6}$. As expected, the two orbital planes are aligned if $\alpha =0$, which corresponds to the coplanar case, where the two stars rotate in the same plane as that of the binary around the black hole.

Finally, we evaluate the sinus of $\psi = |\thetaint-\thetacol|$. It is more practical to compute the true anomalies with the origin set at the pericenter location of the binary center of mass. These angles are denoted by $\thetaintbar$ and $\thetacolbar$ and the relation $\psi = |\thetaintbar-\thetacolbar|$ holds. The tangent of the true anomaly at the plane intersection line is
\be
\begin{split}
\tan \thetaintbar
& = \frac{(\ddj \times \vect{j})\cdot \ey}{(\ddj \times \vect{j})\cdot \ex} \\
& = 2^{1/6} \zeta \, \beta^{-1/2} \left(\frac{a}{\rstar}\right)^{-1/2}, \\
& = 0.011 \pi \, \zeta \, \beta^{-1/2} \a3^{-1/2} \rs^{1/2},
\end{split}
\label{intersection_line}
\ee
where $\zeta = 2(\cos \phi \pm \sqrt{2} \sin \phi)/(2\cos \phi \pm \sqrt{2} \sin \phi)$. This factor obeys $\zeta \approx 1 >0$ for most values of the phase angle, meaning that the intersection line passes through the lower-left and upper-right quadrants with respect the black hole. In this case, the true anomaly can also be safely approximated by $\thetaintbar \approx \pi + \tan \thetaintbar$. The relation $\tan \thetaintbar \approx (\rp/\rtsep)^{1/2}$ shows that the angle $\thetaintbar$ is similar to the true anomaly at which the binary gets separated, computing it at lowest order in $\rp/\rtsep$. The plane intersection line therefore passes most of the time near the point of binary separation. This is unsurprising because the two stars are close together at that location such that their orbital planes are likely to cross in the vicinity. However, there also exists a small range of values for the phase angle $\phi$ for which this line can take any other direction. In particular, $\tan \thetaint = 0$ for $\phi = \mp\arctan(1/\sqrt{2}) \approx \mp 0.2 \pi$ meaning that the intersection line is directed along $\ex$.

To determine the true anomaly at the collision point, we use the fact that it belongs to the interval given by equation \eqref{thetacol_range} to write
\be
\begin{split}
\thetacolbar
& =\thetacol^{\rm mb}-f(\thetacol^{\rm mb}-\thetacol^{\rm mu})-\dtheta/2\\
& \approx \pi -f(\gammamu-\dtheta/2),
\end{split}
\label{collision_angle}
\ee
where the $-\dtheta/2$ term in the first line accounts for the origin of $\thetacolbar$ at the pericenter location of the binary center of mass. The location of the collision point is parametrized by $f$ that satisfies $0\leq f \leq 1$. This parameter takes its lowest value $f=0$ for streams collision involving the most bound stream elements and increases to $f=1$ when the most unbound element of the second stream collides. 

It is then possible to compute the sinus of $\psi$ using
\be
\sin \psi \approx |\cos \thetaintbar \sin \thetacolbar - \sin \thetaintbar \cos \thetacolbar|,
\label{sinpsi}
\ee
in combination with equations \eqref{intersection_line} and \eqref{collision_angle} making use of the relation $|\sin \thetaintbar| = |(\ddj \times \vect{j}) \cdot \ey| / |\ddj \times \vect{j}| \approx 2^{7/6} (\tilde{\zeta}/\chi) \beta^{-1/2} (a/\rstar)^{-1/2}$ where $\tilde{\zeta} = |\cos \phi \pm \sqrt{2} \sin \phi| \leq \sqrt{3}$ is the numerator of $\zeta$ in absolute value. Because $\gammamu > \dtheta/2$, the collision point is the closest to the plane intersection line for $f=0$, which corresponds to a collision involving the most bound stream elements. In this situation, the angle $\psi$ reaches its lowest value $\psi_{\rm min}$. The quantity involved in the condition $\Delta z \leq H$ for streams collision is then also minimum and given by
\be
\begin{split}
i \sin \psi_{\rm min}
& = i \, |\sin \thetaintbar | \\
& \approx 2^{1/3} \tilde{\zeta} \, |\sin \alpha| \left(\frac{\mh}{\mstar}\right)^{-1/3}\\
& =  0.0022 \, \pi^2 (\tilde{\zeta}/\sqrt{3})  \, |\sin \alpha| \m6^{-1/3} \ms^{1/3},
\end{split}
\label{sinpsimin}
\ee
according to equation \eqref{sinpsi}. Remarkably, this term is independent of the binary separation that cancels out in the product. Its numerical value can be injected in equation \eqref{condition_inclination} to evaluate whether a collision takes place between the streams most bound parts. One can already see that the condition $\Delta z < H$ is satisfied for $m=1$ as long as $\sin \alpha \lesssim 0.5$, meaning that the streams can collide with a significant likelihood for homologously expanding streams. However, a much smaller value of $\sin \alpha$ is required to reach this condition if $m=1/4$, implying that collisions are much less likely for streams confined by self-gravity.

\begin{figure}
\includegraphics[width=0.47\textwidth]{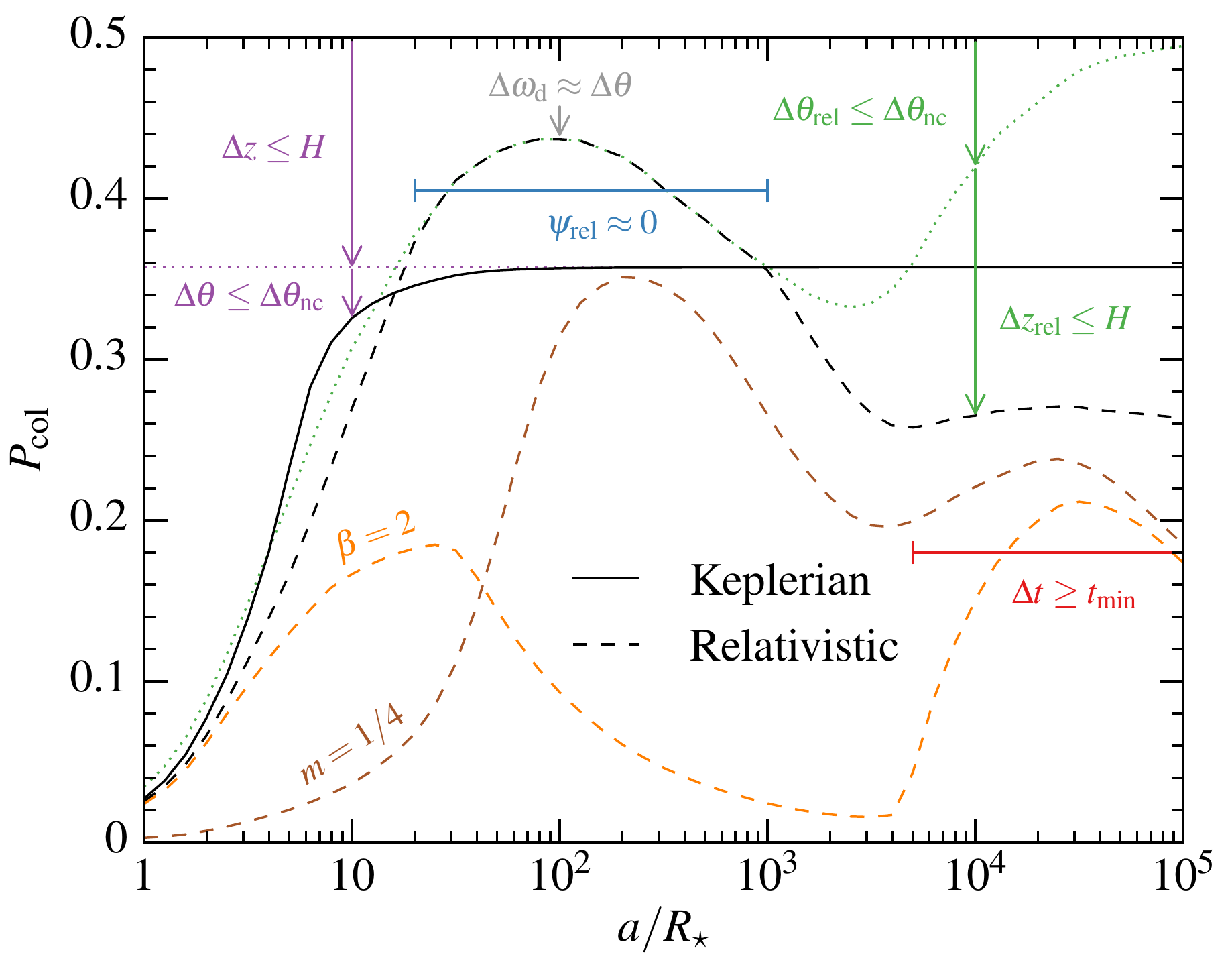}
\caption{Probability of streams collision as a function of binary separation for the Keplerian (black solid line) and relativistic (black dashed line) calculations assuming $\beta=1$ and that the streams expand homologously ($m=1$). The purple arrows represent the reduction of the Keplerian probability from the upper limit of $50\%$ due to the conditions $\Delta z \leq H$ (purple dotted line) and  $\dtheta\leq\dthetanc$. The green arrows show the same for the relativistic probability, for which the reduction results from the conditions $\dthetarel\leq\dthetanc$ (green dotted line) and $\Delta z_{\rm rel} \leq H$. The two segments indicate the ranges of binary separation for which $\psirel \approx 0$ (blue) and $\Delta t \geq \tmin$ (red) while the grey arrow indicates $\domegad \approx \dtheta$. The relativistic probability is also shown for $\beta=2$ (dashed orange line) and assuming that the streams width is confined by self-gravity ($m=1/4$, dashed brown line) for a collision point at $\rcol = 2 \amin$.}
\label{probability}
\end{figure}

A quantitative estimate of the likelihood of streams collision can be obtained from an integral over the random binary angles restricted to a domain where the three conditions for streams collision are all satisfied. In particular, the condition $\Delta z \leq H$ is evaluated for a value of the parameter $f$ that minimizes the vertical offset, which amounts to considering the element of the second stream that is the most likely to collide with the first stream. This choice is legitimate since streams collision occurs if at least one element of each stream collides with each other. The collision probability is then
\be
\pcol=  \frac{1}{8\pi^2} \int_{D_{\rm col}} \cos \alpha \diff \alpha \diff \delta \diff \phi,
\label{collision_probability}
\ee
where $D_{\rm col}$ denotes the domain of integration described above. This integral can be calculated numerically. For simplicity, we compute $\dthetanc$ assuming $\beta_1=\beta_2=1$ (grey region in the left panel of Fig. \ref{planes}) even though the actual penetration factors of the stars can differ from unity. The resulting probability is shown with the solid black line in Fig. \ref{probability} as a function of binary separation for a mass ratio set to $\mh/\mstar = 10^6$. As expected, it is lower than 50\% due to the upper bound imposed by the requirement of a positive pericenter shift. The probability is in addition primarily constrained by the condition $\Delta z \leq H$ that leads to a further decrease to $\pcol \approx 36 \%$ for homologously expansing streams ($m=1$) as indicated by the horizontal purple dotted line. A more drastic decrease happens if the streams are confined by self-gravity ($m=1/4$) that leads a significantly smaller collision probability. This is consistent with the predictions made based on equation \eqref{sinpsimin}, which corresponds to the value of the product $i \sin \psi$ giving  the minimal vertical offset. The fact that this quantity is independent on $a$ also justifies the fact that the probability reduction is the same at all binary separations. The suppression for $a\lesssim 10 \rstar$ is due to the fact that $\dtheta > \dthetanc$ for some values of the binary angles. This is expected from the $\dtheta$-$\dt$ plane of Fig. \ref{planes} because the blue and grey regions intersect if extrapolated downwards to such low binary separations, corresponding to $\dt\lesssim 10^{-5} \tmin$. The likelihood of collisions is therefore high as long as the binary is not close to contact and the streams are homologously expanding.

\subsection{Relativistic corrections}

So far, the conditions for streams collision have been derived assuming perfectly Keplerian trajectories. At pericenter, the gas can in fact approach the gravitational radius of the black hole, implying that relativistic corrections must be accounted for. The main effect to deal with is relativistic apsidal precession that causes a rotation in the direction of motion of the major axis of each star when it passes at pericenter. To include it in our calculation, it is convenient to decompose the angle by which the stars precess into a net and differential component. The net component is the precession angle for a pericenter equal to that of the binary center of mass. It is given by
\be
\domegan \approx \frac{3 \pi G \mh }{\rtdis c^2} \beta \approx 0.064 \pi \, \beta \m6^{2/3} \ms^{1/3} \rs^{-1},
\ee
using the first order approximation of the relativistic precession angle \citetext{equation 10.8 of \citealt{hobson2006}} for a nearly-parabolic orbit. Differential precession is due to the fact that the stars have distinct pericenter distances that make them precess by different amounts. This pericenter variation is induced by the angular momentum kick experienced by the stars during the binary separation. The associated change in penetration factor $\Delta \beta = \beta_2 - \beta_1$ is
\be
\begin{split}
\frac{\Delta \beta}{\beta}
& \approx 4 \, \frac{\vect{j} \cdot \ddj}{\vect{j}^2} \\
& = 2^{2/3} \kappa \, \beta^{1/2}  \left(\frac{\mh}{\mstar}\right)^{-1/3} \left(\frac{a}{\rstar}\right)^{1/2} \\
& \approx 0.87 (\kappa/\sqrt{3}) \, \beta^{1/2} \m6^{-1/3} \ms^{1/3} \a3^{1/2} \rs^{-1/2},
\end{split}
\label{beta_variation}
\ee
where $\kappa=\cos \alpha \sin \delta (\sqrt{2} \cos \phi \pm \sin \phi) + \cos \delta (\sqrt{2} \sin \phi \mp \cos \phi)$ satisfies $-\sqrt{3} \leq \kappa \leq \sqrt{3}$. This yields a differential precession angle of
\be
\begin{split}
\domegad
& \approx \Delta \omega_{\rm n} \frac{\Delta \beta}{\beta}\\
& \approx 0.055 \pi \,  (\kappa/\sqrt{3}) \, \beta^{3/2} \m6^{1/3} \ms^{2/3} \a3^{1/2} \rs^{-3/2},
\label{differential_precession}
\end{split}
\ee
which is by convention positive if the second star precesses more. This precession angle is therefore the largest for wide binaries. Net and differential relativistic precessions modify the pericenter shift angle and the true anomaly of the collision point. The relativistic versions of these quantities are given by
\be
\dthetarel = \dtheta+\domegad,
\label{pericenter_shift_relativistic}
\ee
\be
\thetacolbarrel = \pi -f(\gammamu-\dthetarel/2) + \domegan,
\label{collision_angle_relativistic}
\ee 
which replace equations \eqref{pericenter_shift} and \eqref{collision_angle}, respectively. The location of the intersection line remains the same as in the Keplerian calculation since apsidal precession does not modify the orbital planes of the streams. The relativistic version of the angle $\psi$ is therefore given by $\psirel = |\thetacolrel - \thetaint|$ where $\thetaint$ is still that of equation \eqref{intersection_line} and the resulting vertical offset is denoted $\Delta z_{\rm rel}$.

The pericenter shift is only affected by differential precession. Like $\xi$ in equation \eqref{pericenter_shift}, the function $\kappa$ of equation \eqref{differential_precession} is largely independent of $\alpha$. Adopting $\alpha =0$ without loss of generality, $\dtheta$ and $\domegad$ have different signs only if the effective phase angle obeys $\mp \arctan(1/\sqrt{2}) \leq \phi+\delta \leq \mp \arctan(\sqrt{2})$. In this interval, the pericenter shift $\dtheta$ is negative and positive in the prograde and retrograde case, respectively. This means that, in the prograde case, it is possible to have $\dthetarel\geq0$ while $\dtheta\leq0$, making the relativistic condition of positive pericenter shift slightly more likely than the Keplerian one. The contrary is true in the retrograde case, where the relativistic condition is less likely. However, these opposite contributions cancel out when the equal likelihood of a binary to be prograde and retrograde is accounted for. As a result, the overall probability of $\dthetarel\geq0$ remains of exactly 50\%, like for the Keplerian calculation. More importantly, the increase in pericenter shift can result in $\dthetarel \geq \dthetanc$ that prevents streams collision if the time delay additionally satisfies $\dt \leq \tmin$ (grey region in the $\dtheta$-$\dt$ plane of Fig. \ref{planes}).

The true anomaly of the collision point given by equation \eqref{collision_angle_relativistic} increases with both differential and net precession. This affects the condition $\Delta z_{\rm rel} \leq H$ for streams collision by changing the angle $\psirel$. Interestingly, there exists a range of binary separations for which relativistic precession is such that this angle reaches a minimum of $\psirel = 0$ for specific values of $f$. This means that the collision point is located exactly on the plane intersection line for certain elements of the second stream. For these elements, the condition $\Delta z_{\rm rel} \leq H$ is satisfied despite the inclination of the orbital planes making streams collision more likely. Note that this effect is not to be expected in the Keplerian regime where the minimal value of $\psi$, obtained when $f=0$, is always larger than zero (equation \ref{sinpsimin}).

These relativistic effects modify the probability of streams collision defined in equation \eqref{collision_probability} by changing the domain of integration. This relativistic probability is shown in Fig. \ref{probability} with a black dashed line for $\beta=1$ and a streams width evolving homologously ($m=1$). The condition of positive pericenter shift still imposes an upper bound of $50\%$. In addition, it is mostly constrained by the condition $\dtheta < \dthetanc$ that leads to the reduction indicated by the green dotted line. Its evolution with $a$ originates mostly from the fact that the differential precession angle increases with binary separation as $\domegad \propto a^{1/2}$ (equation \ref{differential_precession}). For $a \lesssim 100 \rstar$, this precession does not affect the pericenter shift since $\domegad \lesssim \dtheta$. This shift is therefore limited to the blue region in the $\dtheta$-$\dt$ plane of Fig. \ref{planes}. Because the size of the grey region decreases with increasing $a \propto \dt^{2/3}$, the condition $\dthetarel \approx \dtheta \leq \dthetanc$ for streams collision becomes more likely, making $\pcol$ larger. The probability reaches a peak at $\pcol \approx 44\%$ but starts to decrease again for $a \gtrsim 100 \rstar$. This is due to $\domegad \gtrsim \dtheta$, which implies that the pericenter shift is not limited to the blue region anymore. Consequently, $\dthetarel>\dthetanc$ for some binary angles that decreases $\pcol$. This decrease stops for $a \gtrsim 5000 \rstar$ (red segment) where the collision probability reaches a plateau at $\pcol\approx 26\%$. This is because $\dt \geq \tmin$ that makes the condition $\dthetarel \leq \dthetanc$ irrelevant, as can be seen from a sharp increase of the green dotted line. On top of this overall evolution, the relativistic probability features two breaks at the edges of the interval $20 \lesssim a/\rstar \lesssim 1000$ (blue segment). Inside this interval, it coincides with the green dotted line, meaning that the condition $\Delta z_{\rm rel} \leq H$ is satisfied for all binary angles. This strong reduction of the vertical offset results from $\psirel\approx 0$, for which, as mentioned above, the location of the collision point determined by relativistic precession coincides with that of the plane intersection line. For binary separations $a \gtrsim 1000\rstar$, this condition becomes more constraining because relativistic precession makes the collision point move away from the plane intersection line, increasing $\psirel$. Fig. \ref{probability} also shows the evolution of the relativistic probability for $\beta=2$ (dashed orange line) and a streams width confined by self-gravity ($m=1/4$, dashed brown line) keeping the other parameters fixed. Increasing the penetration factor leads to a global decrease of the probability since the condition $\dthetarel>\dthetanc$ becomes more constraining owing to an increase of the differential precession angle as $\domegad \propto \beta^{3/2}$ (equation \ref{differential_precession}). A similar decrease is seen for $m=1/4$ because the condition $\Delta z_{\rm rel} \leq H$ for streams collision is less likely to be satisfied owing to the thinner profile of the streams. Nevertheless, the probability is larger than in the Keplerian calculation owing to the reduction of $\psirel$ by apsidal precession. In both cases, the probability also features two peaks that are due to $\domegad \approx \dtheta$ and $\dt\geq \tmin$ at short and wide binary separations, respectively. The probability of streams collision is therefore significant except for near-contact binaries and can be as high as $\pcol \approx 44\%$ in the most favourable configuration.

\begin{table*}
\begin{threeparttable}
\centering
\caption{Parameters of the different models and values of the quantities involved in the conditions for streams collision. The range of ratios $\Delta z/H$ corresponds to the parameter $f$ covering the interval $0\leq f\leq1$.}
\begin{tabular}{@{}lclcrccc@{}}
  \hline
   \multirow{2}{*}{Model} &  \multirow{2}{*}{$\alpha/\pi$} &  \multirow{2}{*}{Rotation} &  \multirow{2}{*}{$\Delta t/\tmin$} &  \multirow{2}{*}{$\Delta \theta/\pi$} & $\Delta z/H$ &  $\Delta z/H$   \\
   & & & & & ($m=1$) & ($m=1/4$) & \\
 \hline
   R$\alpha$0 & 0 & retrograde & 0.09 & 0.0057 & 0 & 0  \\
   P$\alpha$0 & 0 & prograde   & 0.09 & $-$0.0057 & 0 & 0 \\ 
   R$\alpha$0.1 & 0.1 & retrograde & 0.085 & 0.0057 & 0.39-2.4 & 12-78 \\
\hline
\end{tabular}
\label{param}
\end{threeparttable}
\end{table*}

\section{Numerical simulations}
\label{simulations}

We now present numerical simulations carried out in order to demonstrate the validity of the conditions for streams collision derived in Section \ref{conditions}  and to study the hydrodynamics of the interaction. The simulations focus on a double TDE produced by the previous tidal separation of a binary star, for which the initial conditions leading to a collision have been determined in Section \ref{binary_likelihood}. For simplicity, we assume that the gas evolves in a Keplerian gravity and do not investigate the relativistic effects presented above.

\subsection{Setup}

We simulate a double TDE produced by a previous binary separation. The two stars have solar masses and radii. The black hole has a mass $\mh=10^6\msun$ and the binary separation is  $a=1000\rsun$. The numerical simulation is initialized with the binary center of mass located at the tidal separation radius\footnote{A proper treatment of the tidal separation process would require to start the numerical calculation far away from $\rtsep$ where the tidal force on the binary is negligible compared to the gravitational attraction between the stars. Instead, the binary center of mass is initially positioned exactly at the tidal separation radius. This choice is made so that the binary angles defined in Section \ref{binary_likelihood} can be used directly to initialize the calculation, which facilitates the comparison between our analytical predictions and the numerical computation.} and following a parabolic orbit with penetration factor $\beta=2$. This choice of pericenter is made so that the two stellar components enter the tidal disruption radius despite the small angular momentum kick experienced during their separation. The binary angles and the direction of rotation specify the initial positions and velocities of each star that are computed using equations \eqref{position_com}, \eqref{velocity_com}, \eqref{position_kick} and \eqref{velocity_kick}. The yaw and phase angles are kept to $\delta=\phi=0$ for all the models. The first two models have both a pitch angle of $\alpha=0$, implying that the streams produced by the disruptions move on the same plane. These two models only differ by the direction of rotation, which is retrograde for model R$\alpha$0 and prograde for model P$\alpha$0. The time delay between the passage at pericenter of the stars is $\Delta t = 0.09 \, \tmin$ for both models according to equation \eqref{time_delay}. The pericenter shift is of $\dtheta = 0.0057 \pi >0$ for model R$\alpha$0 and $\dtheta = -0.0057 \pi <0$ for model P$\alpha$0 as obtained using equation \eqref{pericenter_shift}. A collision between streams is therefore only expected in the former case. The last model R$\alpha$0.1 also assumes a retrograde rotation. Additionally, it has $\alpha=0.1 \pi$ that implies two different orbital planes for the streams. The pericenter shift is $\dtheta = 0.0057 \pi >0$, that is the same as for model R$\alpha$0 due to the fact that $\delta =0$. Due to the positive $\alpha$, the time delay is slightly reduced to $\dt = 0.085 \, \tmin$. The ratio of vertical offset to streams width induced by the inclination of orbital planes satisfies $12 \leq \Delta z/H \leq 78$ for $m=1/4$ and $0.39 \leq \Delta z/H \leq 2.4$ for $m=1$ according to equations \eqref{condition_inclination}, \eqref{inclination} and \eqref{sinpsi}. This range of values corresponds to different $f$, the lowest one being reached for $f=0$ and the largest for $f=1$. The fact that $\Delta z > H$ for most values of $f$ indicates that streams collision is expected to be weakened for model R$\alpha$0.1 compared to model R$\alpha$0 due to the passage of a large fraction of one stream above the other. Table \ref{param} summarizes the parameters used in each model along with the values of the quantities involved in the condition for streams collision. These quantities are also indicated in the $\dtheta$-$\dt$ plane of Fig. \ref{planes} for models R$\alpha$0 (orange circle) and R$\alpha$0.1 (purple triangle). They are located in the red hatched region implying that, if a collision takes place, the most bound parts of each stream are expected to interact with each other. In the $\psi$-$i$ plane, the purple line corresponds to model R$\alpha$0.1 for the parameter $f$ varying between 0 (leftmost triangle) and 1 (rightmost triangle). The leftmost triangle is below the dashed line indicated that the most bound part of the second stream is expected to collide with the first stream if the width evolves homologously, which is consistent with the fact that $\Delta z/H = 0.39<1$ for $f=0$ and $m=1$

\begin{figure*}
\centering
\includegraphics[width=\textwidth]{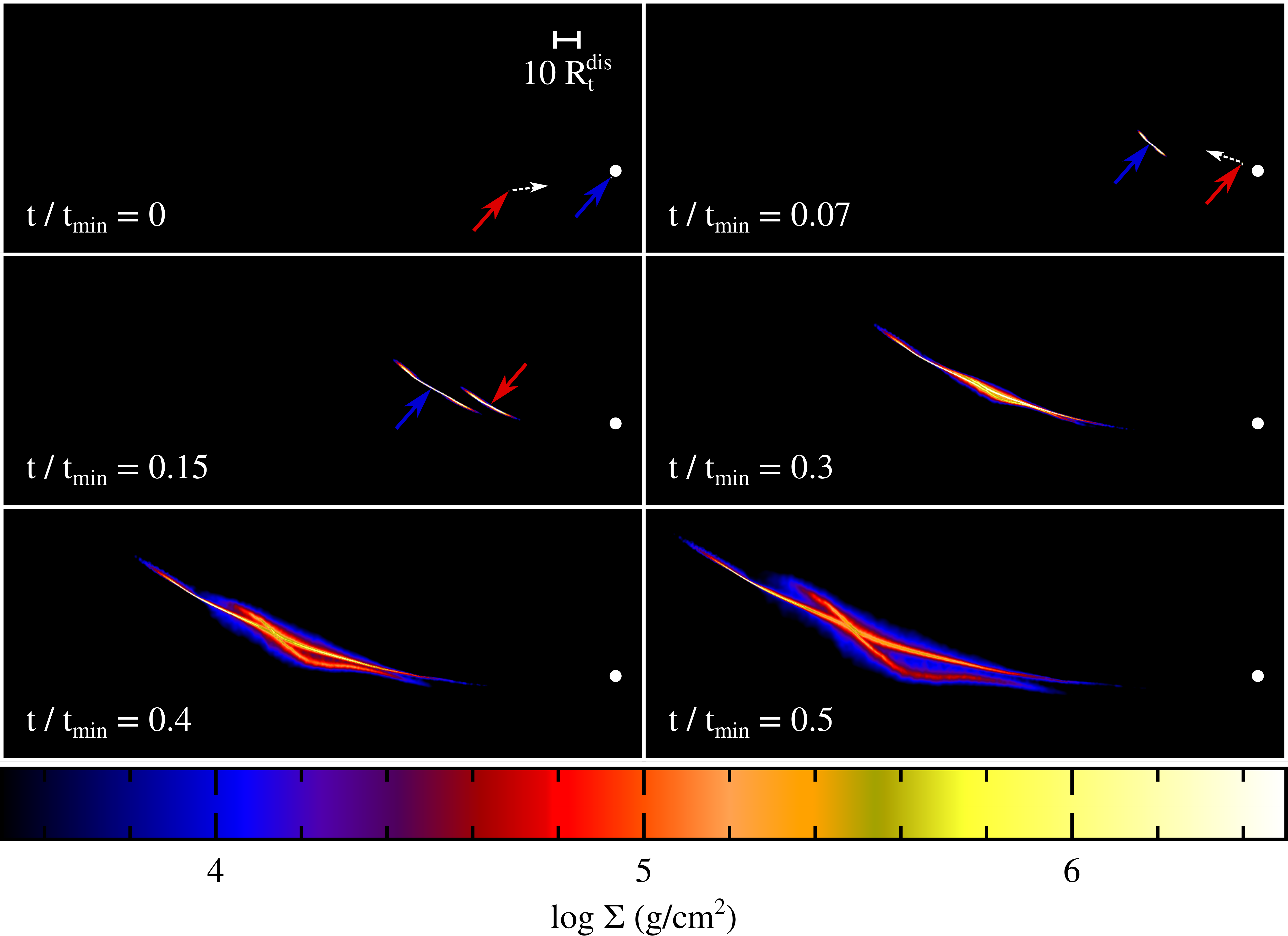}
\caption{Snapshots showing the gas evolution for model R$\alpha$0 at different times $t/\tmin =$ 0, 0.07, 0.15, 0.3, 0.4 and 0.5 during the two stellar disruptions and subsequent evolution of the debris streams. The colours show the gas column density, increasing from blue to yellow as indicated in the colour bar. The black hole is represented by the white dot on the right-hand side of each panel. In the first three panels, the location of the first and second star or stream is indicated with blue and red arrows, respectively. The direction of motion of the second star is shown with a dashed white arrow in the first two panels. All panels use the same scale, indicated by the segment in the first panel that corresponds to ten tidal disruption radii. After the disruptions, the second stream catches up with the first one owing to its positive pericenter shift. This results in a streams collision at $t/\tmin \approx 0.3$ and an associated gas expansion at later times.}
\label{snap-times}
\end{figure*}

The trajectories of the two stars is followed with a three-body calculation performed with the code \textsc{rebound} using the IAS15 integrator \citep{rein2012,rein2015} until the first one reaches a distance of $3\rtdis$ from the black hole.\footnote{\textsc{rebound} can be downloaded freely at \url{http://github.com/hannorein/rebound}.} The positions and velocities of the stars at this point are used to initialize a hydrodynamical simulation, carried out with the SPH code \textsc{phantom} \citep{price2017}. The stars are modelled by polytropic spheres with exponent $\gamma=5/3$ containing $10^5$ SPH particles that we create using the same procedure as \citet{lodato2009}. The black hole gravity is modelled with an external Keplerian potential. Self-gravity is included through a k-D tree algorithm \citep{gafton2011}. Direct summation is used to treat short-range interactions with a critical value of 0.5 in the opening angle criterion. An adiabatic equation of state is assumed for the gas thermodynamical evolution. Shocks are handled with a standard artificial viscosity prescription combined with a switch that strongly reduces its value away from shocks \citep{cullen2010}. Our simulations aim at investigating only the first revolution of the streams around the black hole. For this reason, we remove the SPH particles that come back after the disruptions within a radius of $30 \rtdis$ from the black hole. The size of this region is set such that any particle that falls back enters it before reaching pericenter. Note that this area extends significantly further than the tidal disruption radius, at which gas elements are expected to come back according to angular momentum conservation. This is necessary because streams collision can increase the angular momentum of a fraction of the debris resulting in an increased pericenter distance. Nevertheless, the mass of accreted gas is always negligible within the duration of the simulations.

\begin{figure}
\centering
\includegraphics[width=\columnwidth]{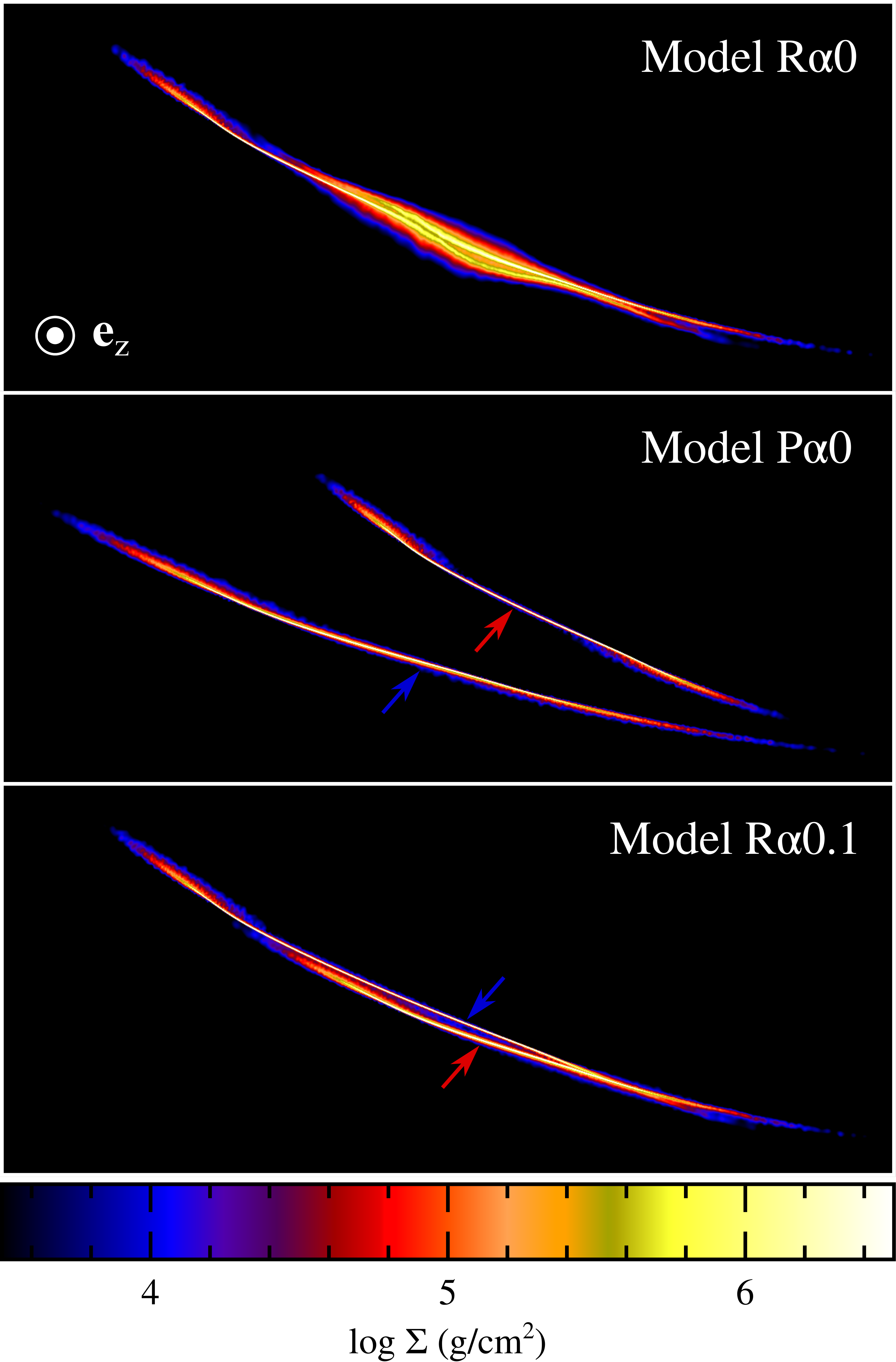}
\caption{Gas distribution at $t/\tmin=0.3$ for models R$\alpha$0 (upper panel), P$\alpha$0 (middle panel) and R$\alpha$0.1 (lower panel) shown along a line of sight perpendicular to the initial binary orbital plane. The colours represent the gas column density, increasing from blue to yellow as indicated in the colour bar. In the two lowermost panels, the first and second stream are indicated by blue and red arrows, respectively. Streams collision occurs around that time for model R$\alpha$0 but is avoided for models P$\alpha$0 and R$\alpha$0.1.}
\label{snap-plane}
\end{figure}

\begin{figure*}
\centering
\includegraphics[width=\textwidth]{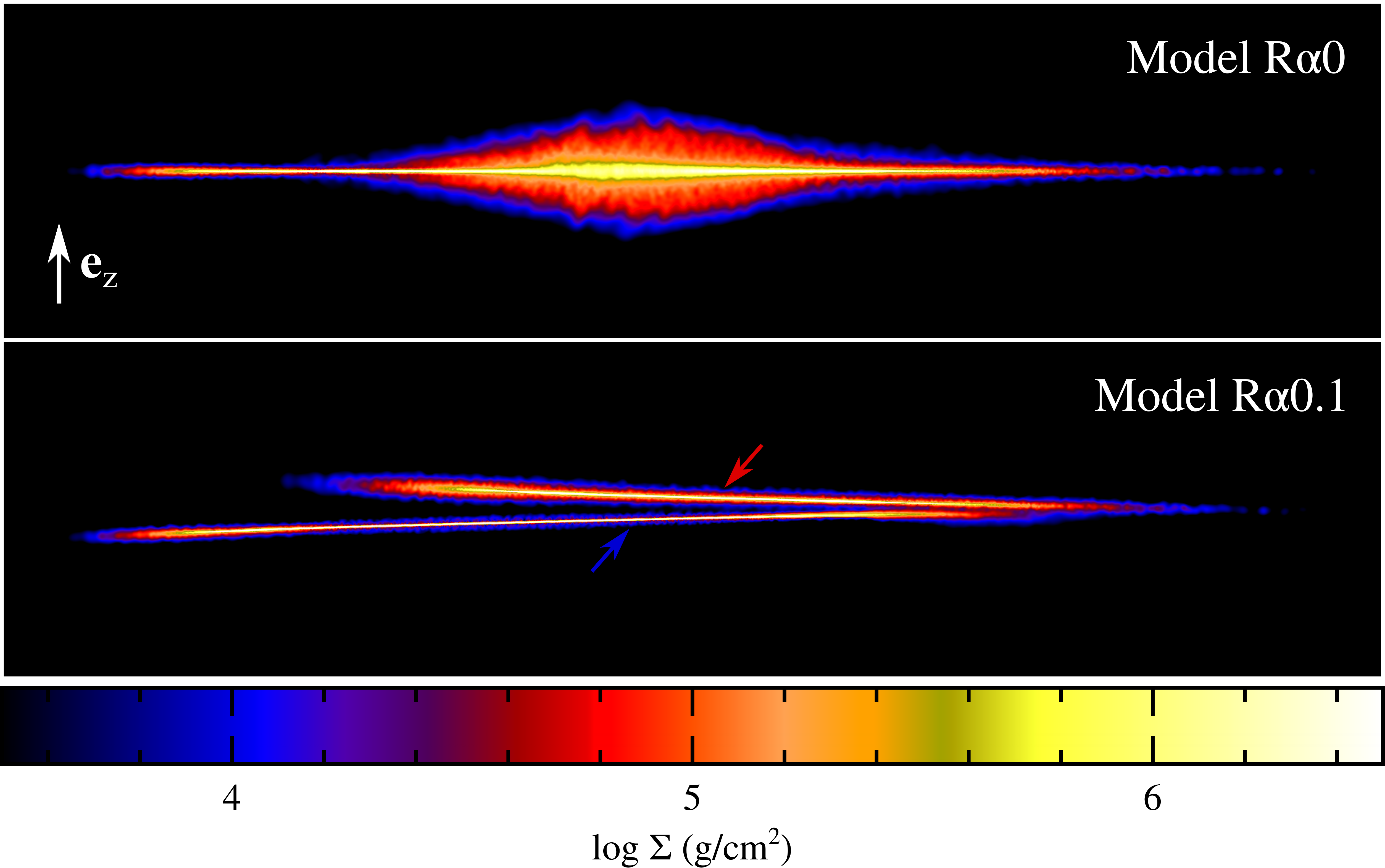}
\caption{Gas distribution at $t/\tmin=0.3$ for models R$\alpha$0 (upper panel) and R$\alpha$0.1 (lower panel) shown along a line of sight parallel to the initial binary orbital plane. The colours represent the gas column density, increasing from blue to yellow as indicated in the colour bar. In the lowermost panel, the first and second stream are indicated by a blue and red arrow, respectively. Most of the streams avoid collision for model R$\alpha$0.1 due to the vertical offset induced by their orbital plane inclination.}
\label{snap-vertical}
\end{figure*}

\subsection{Results}

We now present the results of the SPH simulations for the three models considered.\footnote{Movies of the simulations presented in this paper are available at \url{http://www.tapir.caltech.edu/~bonnerot/double-tdes.html}.} The gas evolution is shown in Fig. \ref{snap-times} for model R$\alpha$0. The black hole is represented by a white dot on the right-hand side of each panel. The blue and red arrows in the first three panels indicate the first and second star or stream, respectively. They are otherwise difficult to identify due to their compactness. The first star is initially closer to the black hole and gets disrupted earlier than the second. The disruptions happen with penetration factors $\beta_1 \approx 1.5$ and $\beta_2 \approx 2.9$ for the first and second star. As expected, these factors differ slightly from that $\beta=2$ of the binary center of mass due to the angular momentum kick given during the separation process. At $t/\tmin = 0.07$, both stars have been disrupted and the streams start their revolution around the black hole. The second stream is still lagging behind the first at $t/\tmin = 0.15$ but is catching up with it owing to the positive pericenter shift. A large fraction of the two streams collide at $t/\tmin \approx 0.3$  leading to an expansion of the gas distribution. At later times, the two streams have partially merged and keep orbiting the black hole, with the bound gas falling back in its vicinity while the unbound part escapes.

The difference in the streams evolution between the models can be understood by looking at Fig. \ref{snap-plane}, which shows the gas distribution at a fixed time $t/\tmin = 0.3$ for models R$\alpha$0 (upper panel), P$\alpha$0 (middle panel) and R$\alpha$0.1 (lower panel). As explained above, the two gas streams collide around that time for model R$\alpha$0. The streams remain instead far apart for model P$\alpha$0 and the collision does not happen. This is a consequence of the negative pericenter shift that prevents the second stream from catching up with the first one. For model R$\alpha$0.1, the second stream is able to catch up thanks to the positive pericenter shift. However, the streams do not strongly collide due to the fact that they evolve on different planes. This situation can be seen more clearly in Fig. \ref{snap-vertical} which shows the gas distribution in the vertical direction for models R$\alpha$0 (upper panel) and R$\alpha$0.1 (lower panel) at the same time of $t/\tmin = 0.3$. The streams collision makes the gas expand vertically for model R$\alpha$0. For model R$\alpha$0.1, the second stream passes above the first stream that prevents most of the debris from interacting. Nevertheless, the most bound parts of the streams undergo a mild encounter due to their smaller offset as can be seen from the lower panel of Fig. \ref{snap-vertical} in the right-hand side. As explained above, this interaction is expected from the fact that $\Delta z/H \approx 0.39 <1$ for the second stream most bound element if the streams evolve homogously. This homologous evolution is caused by the deep disruption of the second star with $\beta_2 \approx 2.9$ that heats the gas at pericenter.

\begin{figure}
\centering
\includegraphics[width=\columnwidth]{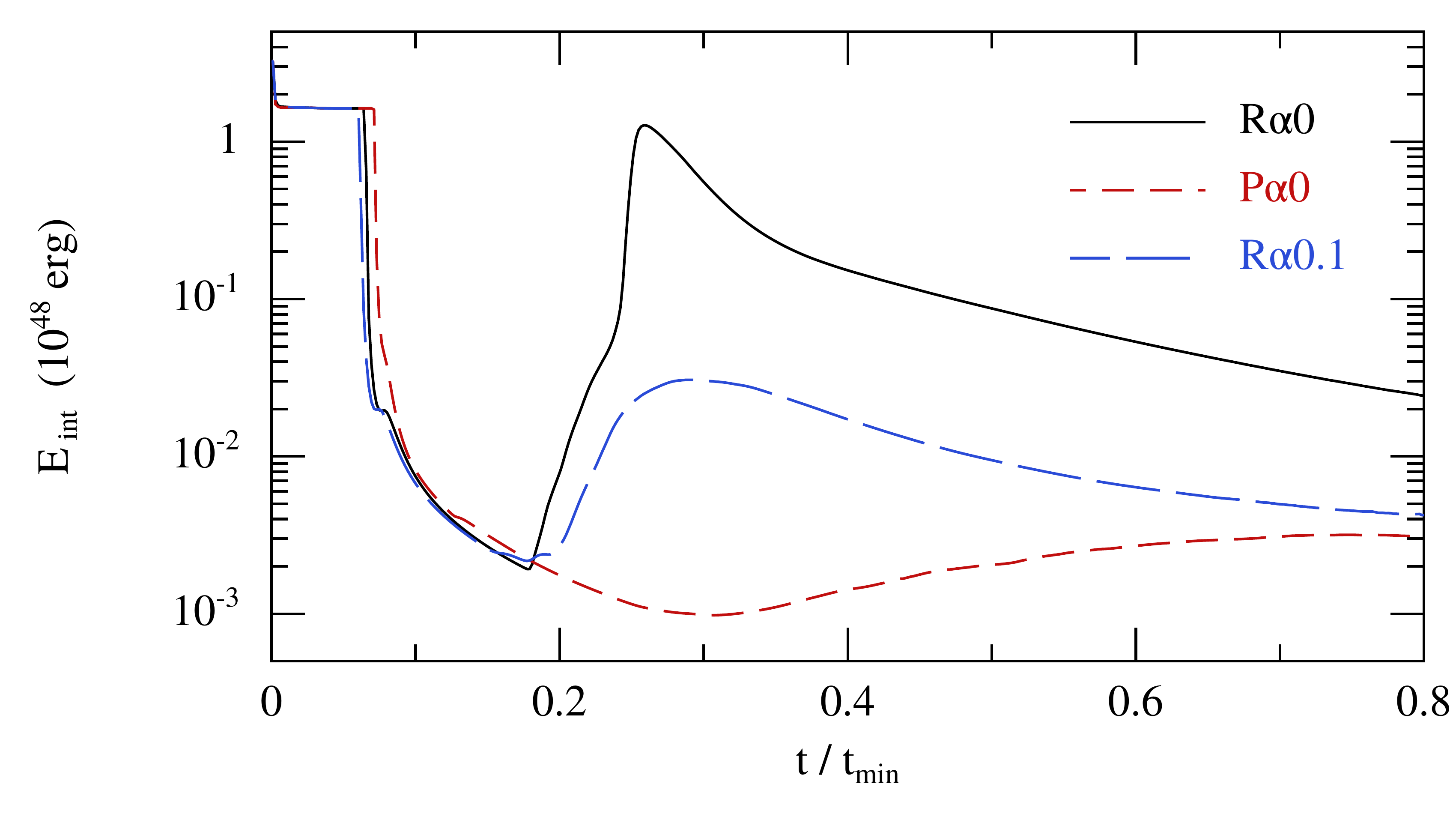}
\caption{Gas internal energy evolution for models R$\alpha$0 (black solid line), P$\alpha$0 (red dashed line) and R$\alpha$0.1 (blue long-dashed line).}
\label{internal-energy}
\end{figure}

A collision between streams is expected to heat the gaseous debris. This effect can be evaluated from Fig. \ref{internal-energy} that shows the internal energy evolution for models R$\alpha$0 (solid black line), P$\alpha$0 (red dashed line) and R$\alpha$0.1 (blue long-dashed line). The early evolution is similar for all models with two sharp drops in thermal energy corresponding to the sequential disruptions. This evolution differs for $t/\tmin \gtrsim 0.2$ where the thermal energy increases for both models R$\alpha$0 and R$\alpha$0 while it keeps decreasing for model P$\alpha$0. The thermal energy increase results from the formation of shocks during the streams collision where a fraction of the gas kinetic energy is dissipated. It peaks at $E_{\rm int} \approx 10^{48} \erg$ for R$\alpha$0 but at a lower value of $E_{\rm int} \approx 10^{46} \erg$ for model R$\alpha$0.1 owing to the smaller amount of gas involved in the collision in the latter case (see lower panel of Fig. \ref{snap-vertical}). At $t/\tmin \gtrsim 0.3$, the thermal energy decreases as the gas expands and cools. For model P$\alpha$0, there is no sharp increase in thermal energy since the streams avoid collision (see middle panel of Fig. \ref{snap-plane}). However, a slow gain in thermal energy can be seen for $t/\tmin \gtrsim 0.3$ until the end of the simulation. This is due to an interaction of the two streams near pericenter. However, the associated shocks are weak since the streams are smoothly joining each other with a small collision angle.

\section{Discussion and conclusion}
\label{discussion}

Several dynamical mechanisms predict TDEs happening at high rates such that two subsequent disruptions may not be independent of each other. In this paper, we investigate the possibility of streams collision resulting from such a double TDE before the debris comes back to pericenter. We start by analytically deriving three conditions for such a collision to happen remaining agnostic about the mechanism at the origin of the TDEs. If the two streams evolve in the same orbital plane, a necessary condition for collision is a positive shift between the pericenter location of the stars, that allows the second stream to catch up with the first one despite its original time delay. However, this pericenter shift must also be lower than a critical value $\dthetanc$ if the time delay is shorter than $\tmin$. Otherwise, collision can be avoided with the second stream passing between the most bound part of the first stream and the black hole. If the orbital planes of the streams are inclined, the collision can also be prevented with one stream passing above the other. In this case, an additional condition for streams collision is that the vertical offset induced by the plane inclination is smaller than the streams width. Using this analytical study, we compute the likelihood of streams collision for a double TDE resulting from a binary separation, treating this process as instantaneous.  The collision probability is significant as long as the binary is not near-contact and reaches $\pcol\approx 44\%$ in the most favourable configuration due to an enhancement induced by relativistic precession. We then perform numerical simulations of a double TDE produced by a binary separation that confirm our analytical conditions for streams collision. If the streams collide, shocks form that result in a sharp increase in thermal energy and a subsequent expansion of the gas distribution.

The fact that streams collision can start before the fallback of the most bound debris at pericenter (equation \ref{minimal_collision_time}) implies that the associated emission represents a precursor to the main flare from TDEs. This early emission could be used to better constrain theoretical models observationally. For example, its detection can help pinpoint the beginning of gas fallback at pericenter in order to get a better handle on the efficiency of disc formation. It is possible to estimate the properties of this signal from our simulations. According to Fig. \ref{internal-energy}, the strongest collision (model R$\alpha$0) leads to a burst of radiation, most likely in the optical band, lasting a few days with a luminosity of $\sim 10^{43} \ergs$ if the internal energy is promptly radiated. This signal could however last up to a few months if the ratio $\dt/\dtheta$ is increased (equation \ref{maximal_collision_time}). According to \citet{mandel2015}, double tidal disruptions resulting from binary separation represent $\sim$10\% of all TDEs. Since the resulting streams have a collision probability of a few tens of percent (see Fig. \ref{probability}), we expect that a precursor is powered through this mechanism in a few percent of TDEs. Furthermore, double disruptions can be produced by other processes that increases the chance of producing this early emission.

The black hole spin has been neglected in our treatment of streams collision. Its main effect is Lense-Thirring precession that causes the angular momentum of the streams to precess around the direction of the black hole spin. Like for relativistic apsidal precession, it is convenient to decompose the Lense-Thirring precession into a net and differential component with associated precession angles given by $\Delta \Omega_{\rm n} \approx 2^{1/2} \pi \, a_{\rm h} \,  \beta^{3/2} (G \mh / \rtdis c^2)^{3/2} \approx  0.0044 \pi \, a_{\rm h} \,  \beta^{3/2} \m6 \ms^{1/2} \rs^{-3/2}$ \citetext{equation 4.220a of  \citealt{merritt2013}} and $\Delta \Omega_{\rm d} = (3/2) \Delta \Omega_{\rm n} \Delta \beta/\beta$. Here, $a_{\rm h}$ denotes the black hole spin parameter. Net precession modifies the position of the plane intersection line. However, the fact that $\Delta \Omega_{\rm n} \ll \domegan$ implies that the impact on the angle $\psirel$ is negligible compared to that induced by apsidal precession. Differential precession changes the inclination angle between the two orbital planes. This modification is nevertheless unable to significantly change the maximal collision probability since it corresponds to $\psirel \approx 0$ as imposed by apsidal precession. This analysis shows that streams collision in double TDEs following binary separation are likely less sensitive to Lense-Thirring precession than stream self-crossing shocks occurring after the gas falls back to pericenter.

Our prediction of a bright TDE precursor associated to streams collision encourages observational attempts to search back for emission, most likely in the optical, during the weeks to months preceding a TDE detection. Such a discovery would unprecedentedly set the scale of times for the phases following the stars disruption, uniquely constraining long standing questions.

\section*{Acknowledgments}

CB and EMR acknowledge the help from NOVA. The research of CB was funded in part by the Gordon and Betty Moore Foundation through Grant GBMF5076. We also thank Yuri Levin, Giuseppe Lodato, Ilya Mandel and Ree'm Sari for insightful discussions. Finally, we acknowledge the use of SPLASH \citep{price2007} for generating the figures of Section \ref{simulations}.




\bibliographystyle{mnras} 
\bibliography{biblio}




\bsp	
\label{lastpage}

\appendix 
\section{Varying the penetration factors}

\label{varying_beta}

In Section \ref{conditions}, the same penetration factor $\beta = 1$ is imposed for the two stars. We now relax this assumption by allowing the two penetration factors to take different values $\beta_1$ and $\beta_2$ for the first and second star. As a result, the eccentricities of equations \eqref{most_bound_eccentricity} and \eqref{most_unbound_eccentricity} differ for the two streams with distinct values $\emin_1$, $\emin_2$, $\emax_1$ and $\emax_2$. For a given stream, the range of debris eccentricities of the elements becomes narrower as the penetration factor increases. This decreases the range of azimuthal angles covered by the debris due to a reduction by a factor $\beta^{1/2}$ of the semi-minor axis of the bound ones and of the impact parameter of the unbound ones. Additionally, a larger penetration factor decreases the time spent by the star near pericenter.

The main effect of varying the penetration factors is to change the two characteristic lines in the $\dtheta$-$\dt$ plane of Fig. \ref{planes}. The parametric function corresponding to an intersection between the most unbound element of the second stream and the most bound element of the first stream is modified to
\be
\dthetanc = \thetacol - \arccos (C_{\rm nc}),
\label{dthetanc_varying}
\ee
\be
C_{\rm nc} = \frac{1}{\emax_2} \left[\frac{\beta_1}{\beta_2}\frac{(1+\emax_2)(1+\emin_1 \cos \thetacol)}{1+\emin_1} -1\right].
\ee
\be
\dtnc = t_1(-\Delta \epsilon,\thetacol,\beta_1) - t_2(\Delta \epsilon,\thetacol - \dthetanc,\beta_2),
\label{dtnc_varying}
\ee
which generalize equations \eqref{dthetanc} and \eqref{dtnc}. Note however that a solution for $\dthetanc$ only exists for a restricted interval of $\thetacol$ if $\beta_1>\beta_2$. This is because the radial position of the first stream element is located close to its pericenter for a collision true anomaly $\thetacol \lesssim \pi/2$. This location cannot be reached by the element of the second stream for any value of the pericenter shift since its pericenter distance is further out. Mathematically, this translates into $C_{\rm nc}>1$ that is outside the allowed domain of equation \eqref{dthetanc_varying}. In equation \eqref{dtnc_varying}, the times needed to reach the intersection point also depend on the penetration factor to account for the modification in time spent near pericenter. This parametric function defined by equations \eqref{dthetanc_varying} and \eqref{dtnc_varying} traces a line in the $\dtheta$-$\dt$ plane of Fig. \ref{planes} for given values of $\beta_1$ and $\beta_2$. Below this line, the streams do not collide for this particular set of penetration factors. For $\beta_1=\beta_2=1$, it is represented by the thick solid black curve that delimits the grey region as explained in Section \ref{conditions}. For values of $\beta_1$ and $\beta_2$ both varying between 1 and 3, the boundary of this region covers the black hatched area. It moves slightly to the left for $\beta_1=\beta_2=3$ (black thin solid line) due to a reduction of the impact parameter of the second stream most unbound debris that allows it to pass between the black hole and the tip of the first stream for a lower $\dtheta$ keeping $\dt$ fixed. The boundary extends further to the left for $\beta_1=1$ and $\beta_2=3$ (black dashed line) due to the reduced time spent by the second stream near pericenter that allows it to catch up with the first stream and avoid collision for a smaller $\dtheta$ at fixed $\dt$. The opposite effect is seen for $\beta_1=3$ and $\beta_2=1$ (black dotted line) where the second stream spends more time near pericenter, resulting in a larger $\dtheta$ required to avoid collision at fixed $\dt$.

Varying the penetration factors also modifies the parametric function corresponding to a collision between the most bound elements of each stream. Equations \eqref{dthetamb} and \eqref{dtmb} are generalized to
\be
\dthetamb =
\begin{cases}
\thetacol-\arccos(C_{\rm mb}), & \thetacol-\dthetamb\leq\pi\\
\thetacol+\arccos(C_{\rm mb})-2\pi, & \thetacol-\dthetamb>\pi\\
\end{cases}
\label{dthetamb_varying}
\ee
\be
C_{\rm mb} = \frac{1}{\emin_2} \left[\frac{\beta_1}{\beta_2}\frac{(1+\emin_2)(1+\emin_1 \cos \thetacol)}{1+\emin_1} -1\right],
\ee
\be
\dtmb = t_1(-\Delta \epsilon,\thetacol,\beta_1) - t_2(-\Delta \epsilon,\thetacol-\dthetamb,\beta_2).
\label{dtmb_varying}
\ee
For a given $\thetacol$, two solutions exist for $\dthetamb$ depending on whether the true anomaly of the second stream element at the collision point satisfies $\thetacol-\dthetamb\leq\pi$ or $\thetacol-\dthetamb>\pi$. These inequalities correspond to a collision while the element of the second stream is moving outwards and inwards, respectively. Not all values of $\thetacol$ lead to a solution for $\beta_1>\beta_2$. In addition to the reason given above that gave $C_{\rm mb}>1$, an additional restriction exists for $\thetacol\approx\pi$, which corresponds to a radial position of the first stream element close to its apocenter that cannot be reached by the second stream element owing to its smaller apocenter distance. It translates into $C_{\rm mb}<-1$ that is again outside the domain allowed by equation \eqref{dthetamb_varying}. As above, the times to collision point of equation \eqref{dtmb_varying} have a dependence on penetration factor that includes the change in time spent near pericenter. The parametric function defined by equations \eqref{dthetamb_varying} and \eqref{dtmb_varying} corresponds to a line in the $\dtheta$-$\dt$ plane of Fig. \ref{planes}. Along this line, the most bound parts of the two streams collide with each other for given values of the penetration factors $\beta_1$ and $\beta_2$. As explained in Section \ref{conditions}, it is shown with a thick solid red line for $\beta_1=\beta_2=1$. For penetration factors both varying between 1 and 3, the line moves slightly upwards to cover the red hatched area. Increasing either $\beta_1$ or $\beta_2$ to 3 while keeping the other at 1 leads to the same dot-dashed red line. For $\beta_1=1$ and $\beta_2=3$, the orbit of the second stream element has a reduced semi-minor axis. As a result, the element of the first stream takes longer to arrive at the collision point while the second stream element reaches it faster. A larger $\dt$ is therefore required for the collision to happen. For $\beta_1=3$ and $\beta_2=1$, the semi-minor axis decreases for the element of the first stream that has the same consequences, implying that $\dt$ must increase by the same amount. For $\beta_1=\beta_2=3$ (thin red line), the combination of these two effects results in an additional increase of $\dt$ needed for the elements to collide.

\end{document}